\documentclass[12pt,preprint]{aastex}
\usepackage{graphicx,color}
\usepackage{color}

\newcommand{\fesc}{$f_{\rm esc}$}

\def\HI{H~{\scriptsize I}}
\def\HII{H~{\scriptsize II}}

\shorttitle{CLUMPING AND THE ESCAPE FRACTION}
\shortauthors{FERNANDEZ \& SHULL}

\begin{document}
\title{The Effect of Galactic Properties on the Escape Fraction of Ionizing Photons}
\author{Elizabeth R. Fernandez, and J. Michael Shull}

\affil{CASA, Department of Astrophysical and Planetary Sciences, University of Colorado,
  389 UCB, Boulder, CO 80309-0389, USA}
\email{
  elizabeth.fernandez@colorado.edu
  michael.shull@colorado.edu
}

\begin{abstract}

The escape fraction, \fesc, of ionizing photons from early galaxies is a crucial parameter
for determining whether the observed galaxies at $z \geq 6$ are able to reionize the high-redshift
intergalactic medium.  Previous attempts to measure \fesc\ have found a wide range of values, 
varying from less than 0.01 to nearly 1.  Rather than finding a single value of $f_{esc}$,  we clarify
through modeling how internal properties of galaxies affect \fesc\ through the density and 
distribution of neutral hydrogen within the galaxy, along with the rate of ionizing photons production.
We find that the escape fraction depends sensitively on the covering factor of clumps, along with 
the density of the clumped and interclump medium.  One must therefore be cautious when dealing 
with an inhomogeneous medium.   
Fewer, high-density clumps lead to a greater escape fraction than more numerous low-density clumps.
When more ionizing photons are produced in a starburst,  \fesc\ increases, as photons escape
more readily from the gas layers.  Large variations in the predicted escape fraction, caused by 
differences in the hydrogen distribution, may explain the large observed  differences in \fesc\  
among galaxies.   Values of \fesc\ must also be consistent with the reionization history.   High-mass 
galaxies alone are unable to reionize the universe, because \fesc\ $> 1$ would be required.   
Small galaxies are needed to achieve reionization,  with greater mean escape fraction in the past. 

\vspace{5cm} 

\end{abstract}

\section{INTRODUCTION}
\label{sec:introduction}

Observations of the cosmic microwave background optical depth made with the {\it{Wilkinson Microwave 
Anisotropy Probe (WMAP)}}
\citep{kogut/etal:2003,spergel/etal:2003,page/etal:2007,spergel/etal:2007,dunkley/etal:2008,komatsu/etal:2008,wmap7}
suggest that the universe was reionized sometime between $6<z<12$.   Because massive stars are efficient producers of 
ultraviolet photons, they are the most likely candidates for the majority of reionization.  However, in order for
early star-forming galaxies to reionize the universe, their ionizing radiation must be able to escape from the halos, in 
which neutral hydrogen (\HI) is the dominant source of Lyman continuum (LyC) opacity.  The escape
fraction, \fesc,  of ionizing photons is a key parameter for starburst galaxies at $z > 6$, which are believed to 
produce the bulk of the photons that reionize the universe 
\citep{robertson/etal:2010,trenti/etal:2010,Bouwens/etal:2010}.  

The predicted values of escape fraction span a large range from $0.01 \lesssim f_{\rm esc} < 1$, derived from a variety of
theoretical and observational studies of varying complexity. Various properties of the host galaxy, its stars, or its environment are thought to affect the number of ionizing photons that escape into the intergalactic medium (IGM).   
For example, \citet{ricotti/shull:2000} studied \fesc\ in spherical halos using a Str\"omgren approach.   \citet{wood/loeb:2000} assumed an isothermal, exponential disk galaxy and followed an ionization front through the galaxy using three-dimensional Monte Carlo radiative transfer.  Both 
\citet{wood/loeb:2000} and \citet{ricotti/shull:2000} state that \fesc\ varies greatly, from $<0.01$ to $1$, depending on galaxy mass, with larger galaxies giving smaller values of \fesc.  A similar dependence with galaxy mass is also seen by the simulations of \citet{yajima/etal:2010}, because larger galaxies tend to have star formation buried within dense hydrogen clouds, while smaller galaxies often had clearer paths for escaping ionizing radiation.    \citet{gnedin/etal:2008}, on the other
hand, ran a high resolution N-body simulation with adaptive-mesh refinement in a cosmological context.  Contrary to \citet{ricotti/shull:2000}, \citet{wood/loeb:2000} and \citet{yajima/etal:2010}, they state that lower-mass galaxies have significantly smaller
 \fesc, as the result of a declining star formation rate.  In addition, above a critical halo mass, \fesc\  does not change by much.  
The model of  \citet{gnedin/etal:2008} allowed for the star formation rate to increase with the mass of the galaxy at a higher rate than a linear proportionality would allow.  The larger galaxies also tended to have star formation occurring in the outskirts of the galaxy, which made it easier for ionizing photons to escape.  Their model included a distribution of gas within the galaxy, which created free sight-lines out of the galaxy.  
\citet{wise/cen:2008} used adaptive mesh hydrodynamical simulations on dwarf galaxies.  Even though their simulations covered a different mass range than the larger galaxies studied by \citet{gnedin/etal:2008}, they found much higher value of \fesc\ than would be expected from extrapolating results from \citet{gnedin/etal:2008} to lower masses.  \citet{wise/cen:2008} attribute this difference to the irregular morphology of their dwarf galaxies with a turbulent and clumpy interstellar medium (ISM), allowing for large values of \fesc.

Others have also looked at how the shape and morphology of the galaxy can affect \fesc.  \citet{dove/shull:1994}, using a Str\"omgren model, studied how \fesc\ varies with various \HI\  disk density distributions.  In addition, many authors have found that superbubbles and shells can trap radiation until blowout, seen in analytical models of \citet{dove/shull/ferrara:2000} as well as in hydrodynamical simulations of \citet{fujita/etal:2003}.  The analytical model by \citet{clark/oey:2002} showed that high star formation rates can raise the porosity of the ISM and thereby increase \fesc.  In addition to bubbles and structure caused from supernovae, galaxies can have a clumpy ISM whose inhomogeneities affect \fesc.  For example, dense clumps could reduce \fesc\  \citep{dove/shull/ferrara:2000}.  On the other hand,
\citet{boisse:1990}, \citet{hobson/scheuer:1993}, \citet{witt/gordon:1996},
and \citet{wood/loeb:2000} all found that clumps in a randomly distributed medium cause \fesc\ to rise, while 
\citet{ciardi/etal:2002} found that the effects of clumps depend on the ionization rate.

A host of other galaxy parameters have been tested analytically and with simulations.  Increasing the baryon mass fraction lowers \fesc\ for smaller halos, but increases it at masses greater than $10^8 \: M_{\sun}$
\citep{wise/cen:2008}.  Star formation history changes the amount of ionizing photons and neutral hydrogen, causing  
\fesc\ to vary from $0.12$ to $0.20$ for coeval star formation and from $0.04$ to $0.10$ for a time-distributed starburst 
\citep{dove/shull/ferrara:2000}.   Other galactic quantities, such as spin \citep{wise/cen:2008} or dust 
content \citep{gnedin/etal:2008}, do not seem to affect the escape fraction.  

Observations have also been used to constrain \fesc, especially at $z\lesssim3$.  Searches for escaping Lyman continuum radiation at redshifts $z \lesssim 1-2$ have found escape fractions of at most a few percent {\citep{bland/maloney:2002,bridge:2010,cowie/etal:2009, tumlinson/etal:1999,deharveng:2001,grimes/etal:2007,grimes:2009, heckman/etal:2001,Leitherer/etal:1995, malkan:2003, Siana/etal:2007}.  
\citet{hurwitz/etal:1997} saw large variations in the escape fraction, and \citet{hoopes:2007} and \citet{bergvall/etal:2006} saw a relatively high escape fraction of $10\%$.  \citet{ferguson/etal:2001} observed  \fesc\ $\approx 0.2$ at $z\approx1$.  \citet{hanish/etal:2010} do not see a difference in \fesc\ between starbursts and normal galaxies.  \citet{siana/etal:2010} also found low escape fractions at $z \approx 1.3$ and showed that no more than $8\%$ of galaxies at this redshift can have $f_{\rm esc,rel} > 0.5$.  Note that  $f_{\rm esc,rel}$, which the authors use to compare their results to other surveys, is defined as the ratio of escaping LyC photons to escaping 1500 $\AA$
photons.  In our own Galaxy, \citet{bland/maloney:1999} and \citet{putman/etal:2003} found an escape fraction of only a few percent.  Observations using $\gamma$-ray bursts \citep{chen:2007} show \fesc\ $\approx 0.02$ at $z \approx 2$.  At higher redshift ($z \approx 3$), \fesc\ seems to vary drastically from galaxy to galaxy \citep{shapley/etal:2006,Iwata/etal:2009, vanzella:2010}, with a few galaxies having very large escape fractions.  Some studies have found low values of the escape fraction at $z \approx 3$ \citep{fernandez-soto/etal:2003,Giallongo/etal:2002,heckman/etal:2001, inoue/etal:2005, wyithe:2010}, while others  have found significant LyC leakage \citep{Steidel:2002,shapley/etal:2006}.  This large variation from galaxy to galaxy suggests a dependence on viewing angle and could indicate the patchiness and structure of neutral hydrogen within the galaxy \citep{bergvall/etal:2006,deharveng:2001,grimes/etal:2007,heckman/etal:2001}.   From these observations, one infers that the fundamental properties of the galaxies change with time or that \fesc\ increases with increasing redshift \citep{bridge:2010,cowie/etal:2009,inoue/etal:2006,Iwata/etal:2009, Siana/etal:2007}.  \citet{Bouwens/etal:2010} looked at the blue color of high redshift ($z \approx 7$) galaxies and
argued that the nebular component must be reduced.  This would suggest a much larger escape fraction in the past. 

The minimum mass of galaxy formation can also put limitations on \fesc.  Observations of Ly$\alpha$ absorption toward high-redshift quasars, combined with the UV luminosity function of galaxies, can limit \fesc\ from a redshift of $5.5$ to $6$, with \fesc\ $ \sim$ 0.20--0.45 if the halos producing these photons are larger than $10^{10} \: M_{\sun}$.  This can decrease to \fesc\ $\sim$ 0.05--0.1 if halos down to $10^8 \: M_{\sun}$ are included as sources of escaping ionizing photons \citep{srbinovsky/wyithe:2008}.  

It is clear that many factors can affect \fesc,  and the problem is quite complicated.   Cosmological simulations that predict the escape fraction provide a more accurate estimate for \fesc.  However, many parameters of the galaxy change at once, and it becomes difficult to understand how a single parameter can affect the escape fraction.  In addition, trends may be difficult to understand because of the manner in which some physical processes are included or neglected.  Analytic models can show clearer trends, even though they may be over-simplified and miss important physics.   Therefore, rather than predicting a quantitative value for \fesc}, we seek to understand how properties of galaxies and their internal structure affect the escape fraction.  Because our model is simplified, the values of \fesc\
are not exact, but rather illustrate trends caused by various galactic properties.  In section \ref{sec:Method}, we explain our method of tracing photons that escape the galaxy.  In section \ref{sec:Results}, we explain our results
and compare our results to previous literature in section \ref{sec:Lit}.  In section \ref{sec:reion}, we consider constraints from reionization and we conclude in section \ref{sec:Conclusions}.  Throughout, we use the cosmological parameters from WMAP-7  \citep{wmap7}.

\newpage

\section{METHODOLOGY}
\label{sec:Method}

\subsection{Properties of the Galaxy}

We use an exponential hyperbolic secant profile \citep{spitzer:1942} to
describe the density of an isothermal disk in a halo of mass $M_{\rm halo}$:
\begin{equation}
n_H(Z) = n_0 \exp[-r/r_{h}]\: \rm{sech}^2\left(\frac{Z}{z_0}\right),
\label{eq:densitysech}
\end{equation}
\citep{spitzer:1942} where $n_0$ is the number density of hydrogen at the
center of the galaxy, $Z$ is the height above the galaxy mid-plane, and $r_{h}$ is the scale radius:
\begin{equation}
r_{h} = \frac{j_d \lambda }{\sqrt{2}m_d}  r_{\rm vir}
\end{equation}
\citep{mo/etal:1998}.  The parameter $j_d$ is the fraction of the halo's
angular momentum in the
disk, $\lambda$ is the spin parameter, $m_d$ is the fraction of the halo in the disk ($m_d=\Omega_b/\Omega_m$), and $r_{vir}$ is the virial radius.  As in \citet{wood/loeb:2000}, we assume $j_d/m_d = 1$ and $\lambda = 0.05$.  The virial radius is  
\begin{equation}
r_{vir}=(0.76 {\rm kpc})  \left(\frac{M_{halo}}{10^8 M_{\sun}
  h^{-1}}\right)^{1/3}\left(\frac{\Omega_m}{\Omega(z_f)}\frac{\Delta_c}{200}\right)^{-1/3}
     \left(\frac{1+z_f}{10}\right)^{-1} h^{-1} \: , 
\label{eq:rvir}
\end{equation}
\citep{navarro/etal:1997} where $\Delta_c = 18 \pi^2 +82d-39d^2$ and
$d=\Omega_{zf}-1$.  $\Omega_{zf}$ is the local value of $\Omega_m$ at the redshift of galaxy formation, $z_f$.  
The dependence of the virial radius on $z_f$ will affect the density of the disk, with smaller disks of higher density 
forming earlier.    
\noindent The disk scale height, $z_0$, is given by 
\begin{equation}
z_0 = \left(\frac{\langle v^2 \rangle}{2\pi G \rho_0}\right)^{1/2} = \left(\frac{M_{halo}}{2\pi \rho_0 r_{vir}}\right)^{1/2},
\end{equation}
where $\langle v^2 \rangle$ is the mean square of the velocity and $\rho_0$ is the central
density.  In a real galaxy, there will be non-thermal motions of the gas.  However, to simplify the calculations, we assume that the gas is virialized, feels the gravity of the disk, and therefore follows the relation $\langle v^2 \rangle = G M_{halo} / r_{vir}$.  However, radiative cooling can cause the disk to be thinner than this.  The central density is solved for in a self-consistent way after the halo mass and the redshift of formation are specified.  We use $15 r_{h}$ and $2z_0$ as the limits of the radius and height of the disk, respectively.  The structure of real galaxies is more complicated.  The addition of stars in the halo will change the gas distribution through feedback, heating, and gravitational effects.  For purposes of simplicity, we ignore these effects.

The mass of the disk (stars and gas) is taken as $M_{\rm disk} = m_d \, M_{\rm halo}$,
where $m_d$ is the fraction of matter that is incorporated into the disk.  The upper limit of $m_d$ is 
$\Omega_b / \Omega_m$.  The mass of the stars within the disk is $M_* = M_{\rm disk} f_* $, where $f_*$ is the star formation efficiency, which describes the fraction of baryons that form into stars.\footnote{The star formation efficiency is defined by the fraction of baryons in stars at any given time.  Therefore, the escape fraction calculated is the escape fraction at that point in the galaxy's lifetime.  The total escape fraction of the galaxy will depend on the lifetime of the burst and the star formation history, which can be obtained by integrating $f_*$ over the duration of the burst, taking into account the propagation of the ionizing front within the galaxy.}  
The remainder of the mass of the disk is in gas, distributed according to equation \ref{eq:densitysech}
with gas temperature of $10^4$ K.  

The number of ionizing photons is related to $f_*$, considering either Population III (metal-free) or
Population II (metal-poor, $Z = 0.02 Z_\sun$) stars.  The total number
of ionizing photons per second from the entire stellar population, $Q_{\rm pop}$ is
\begin{equation}
Q_{pop} = \frac{\int_{m_1}^{m_2}\overline{Q}_H(m) f(m) dm}{\int_{m_1}^{m_2} m f(m) dm}
\times M_* \; ,
\end{equation}
where $m$ is the mass of the star, and $m_1$ and $m_2$ are the upper and lower
mass limits of the mass spectrum, given by
$f(m)$.  For a less massive distribution of stars, we use the Salpeter initial mass spectrum \citep{salpeter:1955}:
\begin{equation}
f(m) \propto m^{-2.35},
\label{eq:salpeter}
\end{equation}
with $m_1=0.4 \;  M_\sun$ and $m_2 = 150 \;  M_\sun$.
The Larson initial mass spectrum illustrates a case with heavier stars \citep{larson:1998}:
\begin{equation}
f(m)\propto m^{-1}\left(1+\frac{m}{m_c}\right)^{-1.35},
\end{equation}
with $m_1 = 1 \;  M_\sun$, $m_2=500 \; M_\sun$, and $m_c = 250 \; M_\sun$ for
Population III stars and $m_1 = 1 \;  M_\sun$,
 $m_2=150 \;  M_\sun$, and $m_c = 50 \;  M_\sun$ for Population II stars.  We
define $\overline{Q}_H$  as
the number of ionizing photons emitted per second per star, averaged over
the star's lifetime.  For Population III stars of mass
parameter $x \equiv \log_{10}(m/M_\sun)$, this is
\begin{eqnarray}
   \log_{10}\left[\overline{Q}_H/{\rm s^{-1}}\right]
      &=&\left\{
     \begin{array}{ll}
     43.61 + 4.90x - 0.83x^2 & 9-500~M_\sun \; ,\\
      39.29 + 8.55x & 5-9~M_\sun \; ,\\
      0 & \rm {otherwise} \; ,
      \end{array}\right.
\end{eqnarray}
and for Population II stars, 
\begin{eqnarray}
   \log_{10}\left[\overline{Q}_H/{\rm s^{-1}}\right]
      &=&\left\{
      \begin{array}{ll}
        27.80 + 30.68x - 14.80x^2
        + 2.50x^3 & \geq 5 M_\sun\\
        0 & \rm {otherwise} \; ,
      \end{array}\right.
\end{eqnarray}
as given in
Table~6 of \citet{schaerer:2002}.

\subsection{Calculating the Escape Fraction}

We place the stars at the center of the galaxy.  An ionized H~{\footnotesize{II}} region develops around the stars, where the number of ionizing photons emitted per second by the stellar population, $Q_{\rm pop}$, is balanced by recombinations, such that
\begin{equation} 
Q_{pop} = \frac{4}{3} \pi r_s^3 n_H^2 \alpha_B(T).
\end{equation}
Here, $\alpha_B$ is the case-B recombination rate coefficient of hydrogen and $T$ is
the temperature of the gas (we assume $T=10^4K$).  The radius
of this \HII\  region, called the Str\"omgren radius, is
\begin{equation} 
r_s= \left(\frac{3 Q_{\rm pop}}{4 \pi n_H^2 \alpha_B}\right)^{1/3}.
\end{equation}
This radius is simple to evaluate in the case of a uniform medium, but if we are concerned with clumps and a disk with a density profile, the density will be changing with location.  Although we are calculating \fesc\ at a moment in time, in reality
the \HII\  region is not static, and the ionization front will propagate at a flux-limited speed.

We assume that all stars are placed at the center of the galaxy.  In reality, star formation will be distributed throughout the galaxy.  If stars are closer to the edge of the galaxy, their photons will have less hydrogen to traverse, and hence will escape more easily.  Therefore, the results we present will be lower limits of the escape fraction.   We integrate along the path length that a photon takes in order to escape the galaxy, following the formalism in \citet{dove/shull:1994}. We can then calculate the escape fraction of ionizing photons along each ray emanating from the center of the galaxy by
equating the number of ionizing photons to the number of hydrogen atoms across its path.  If there are more photons than hydrogen atoms, the ray can break out of the disk; otherwise, no photons escape and the escape fraction is zero.  The escape fraction along a path, $\eta$, thus depends on the amount of hydrogen the ray transverses, which depends on its angle $\theta$, measured from the axis perpendicular to the disk:
\begin{equation} 
    \eta(\theta) = 1 -\frac{4 \pi \alpha_B}{Q_{pop}}\int^{\infty}_0 n_H^2(Z)r^2  dr \; .
\end{equation}
Photons are more likely to escape out of the top and bottom of the disk,
rather than the sides, because there is less path length to traverse.  This
creates a critical angle, beyond which photons no longer will escape the
galaxy.  The total escape fraction, \fesc, is then found by integrating over all angles
$\theta$ and the solid angle $\Omega$:
\begin{eqnarray} 
f_{\rm esc}(Q_{\rm pop})  &=& \int \int \frac{\eta(\theta) }{4\pi} d\theta d\Omega\\
                                &=& \int \frac{1}{2}\eta(\theta)\: \sin(\theta)\: d\theta \; .
\end{eqnarray}
We take into account the whole disk (top and bottom) so that an \fesc\ of $1$ means that all photons produced are escaping into the IGM.

\subsection{Adding Clumps}
A medium with clumps can be described with the density contrast $C = n_{c}/n_{ic}$ between the
clumps (density $n_{c}$), and the interclump medium (density $n_{ic}$). The percentage of volume taken
up by the clumps is described by the volume filling factor $f_{V}$.  We randomly distribute clumps throughout the
galaxy.  We define $n_{\rm mean}$ as the density the medium would have if it was not clumpy, given
by equation \ref{eq:densitysech}.  The density at each point is given by
\begin{equation} 
    n_{c}=\frac{n_{\rm mean}}{f_{V}+(1-f_{V})/C}
\label{eq:clump}
\end{equation}
if the point is in a clump and 
\begin{equation} 
     n_{ic}=\frac{n_{\rm mean}}{f_{V}(C-1)+1}
\label{eq:ic}
\end{equation}
if the point is not in a clump, similar to \citep{wood/loeb:2000}.  In this way, the galaxy retains the same interstellar gas
mass, independent of $f_{V}$ and $C$.  As $C$ increases, the density of the clumps increases as the density of the 
non-clumped medium falls. Similarly, if $f_V$ is larger, more of the medium is contained in less
dense clumps.  We trace photons on their path through the galaxy and
track whether or not they encounter a clump.  At each step on the path out of the galaxy, a random number is generated.  A clump exists if this random number is less than the volume filling factor.  As the filling factor increases, clumps can merge, forming larger, arbitrarily shaped clumps.  Counting the number of photons exiting the galaxy then leads to \fesc.   The covering factor is computed by counting how many clumps intersect a ray as it travels out of the galaxy.


\section{RESULTS}
\label{sec:Results}

\subsection{Properties of the Clumps}

\begin{figure}[t]
\centering \noindent
\includegraphics[width=120mm]{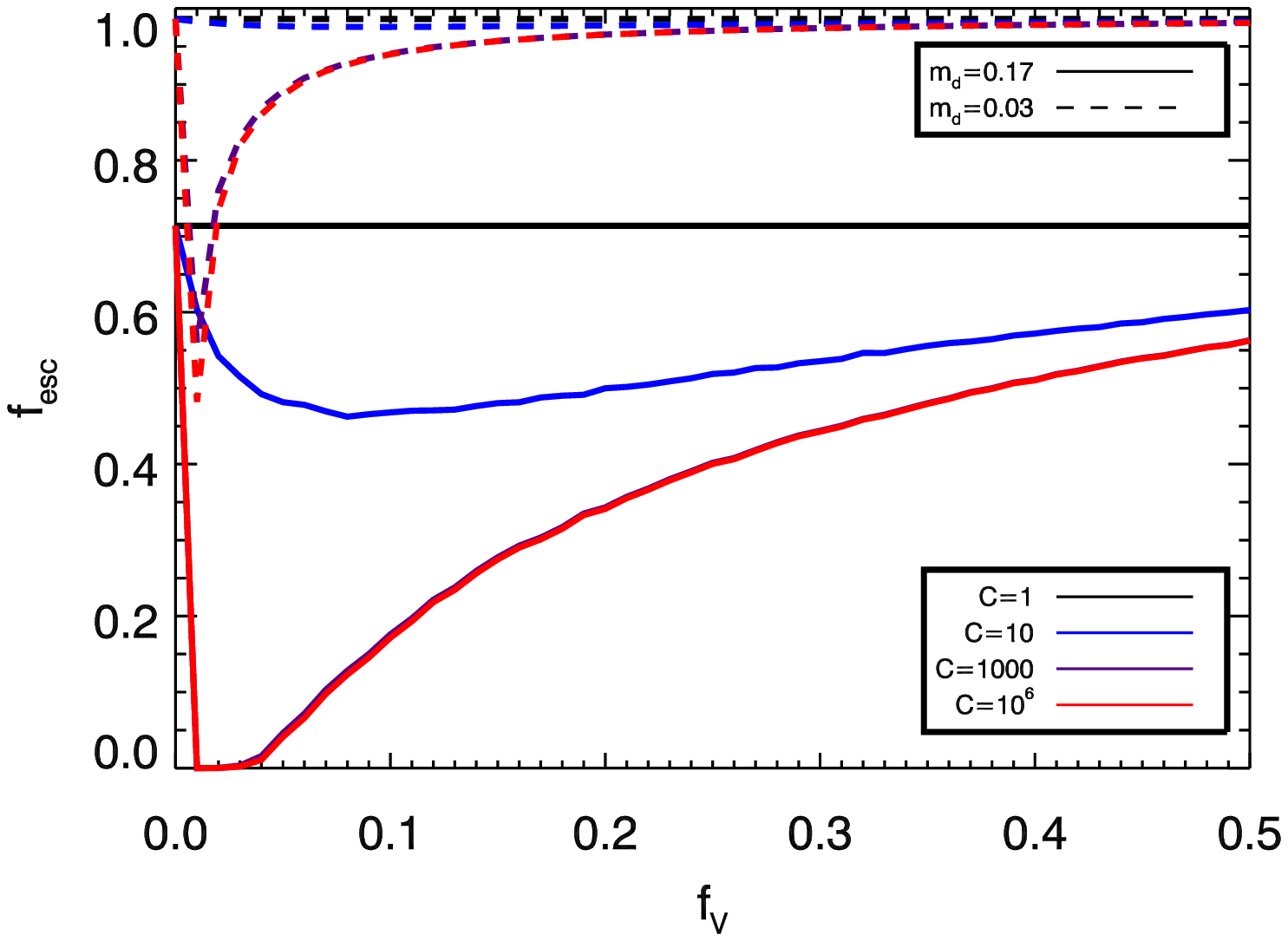}
\includegraphics[width=120mm]{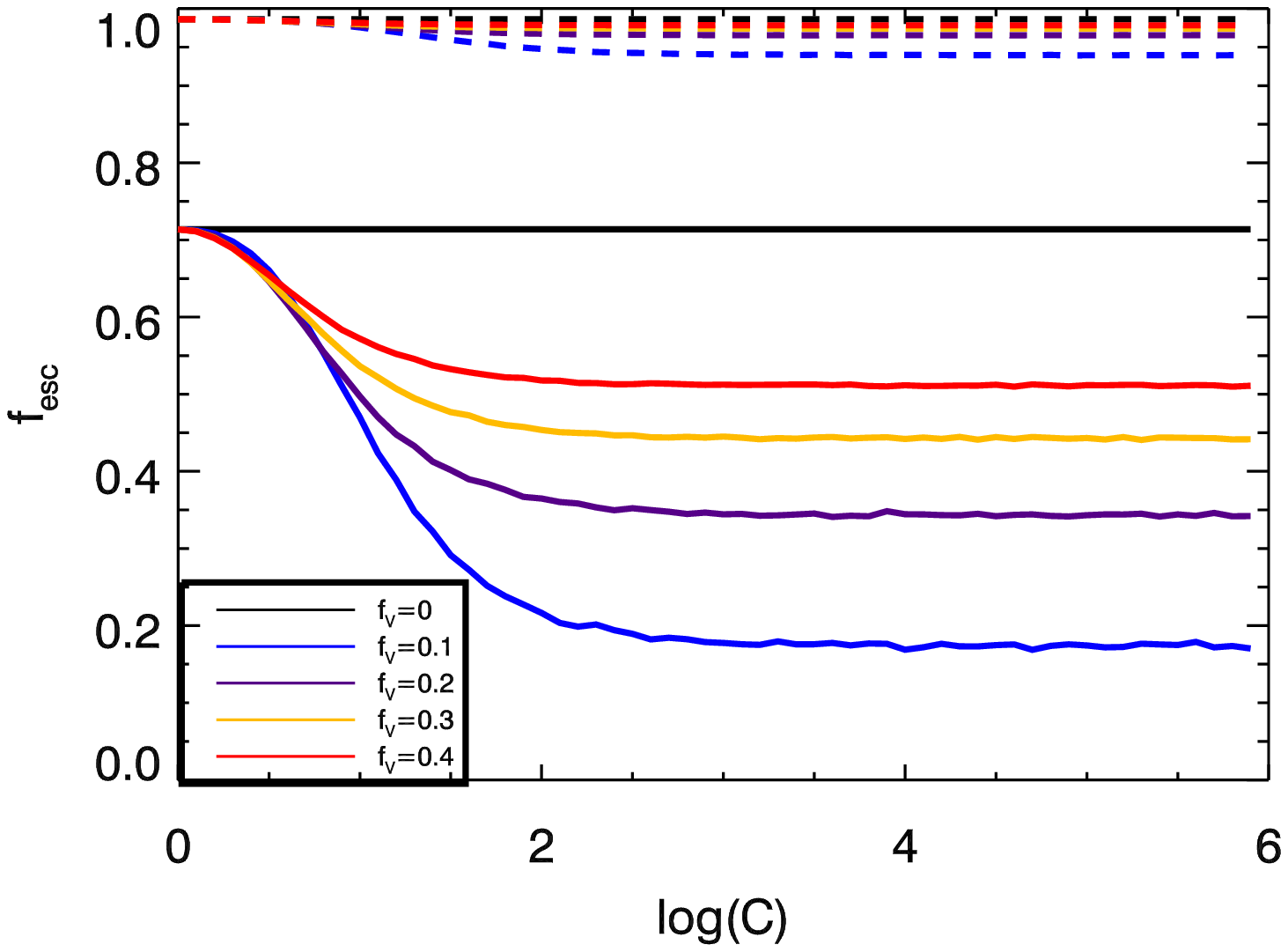}
\caption{
Escape fraction of ionizing photons out of the disk as a function of the clump volume filling factor $f_V$ for various
values of the clumping factor $C$ ({\it{top panel}}) and  $\log(C)$ for various values of $f_V$ ({\it{bottom panel}}).  
Shown for a $10^9 M_\sun$ halo at $z_f=10$, with $f_* = 0.5$ and Population III stars with a Larson mass spectrum.  The fraction of matter incorporated into the disk, $m_d$, scales with the escape fraction.
}
\label{fig:varyff}
\end{figure}

In the first calculations, we placed Population~III stars with a Larson mass spectrum and a star formation efficiency $f_* = 0.5$ in a halo of $M_{\rm halo} = 10^9 M_\sun$, with a redshift of formation of $z_f = 10$.  The clumps have diameter 
$10^{17}$~cm, unless otherwise stated.   The top panel of Figure \ref{fig:varyff} shows \fesc\ as a function of $f_V$ for various values of $C$.  The case with no clumps is equivalent to $C=1$.  As clumps are introduced, \fesc\
quickly falls, but rises again as $f_V$ rises.  This is because the clumps become less dense (since more of the medium is in clumps and the mass of the galaxy must be kept constant).  In addition, \fesc\ drops as $C$ increases, showing that
denser clumps with a less dense interclump medium stops more ionizing radiation than a more evenly distributed medium.  The clumps are small enough that essentially every ray traversing the galaxy encounters one of these very dense clumps and is diminished.

In the bottom panel of Figure \ref{fig:varyff}, the same population of stars is shown for various
values of $f_V$ as a function of $\log(C)$.  As $C$ increases, \fesc\ becomes low for small
values of $f_V$.  Again, this is because of a few very dense clumps that stop essentially all radiation.  
As $f_V$ increases, more of the medium is in clumps, and therefore the density of the clumps
decreases.  The combined effect is an increase of \fesc.  The solid black line shows the case with no clumps, or when $f_V=0$.  For $f_V=0$, \fesc\ equals the case with no clumps ($C=1$), as it should.  Above $C \sim 10-100$,
increasing $C$ no longer affects \fesc.


\begin{figure}[t]
\centering \noindent
\includegraphics[width=80mm]{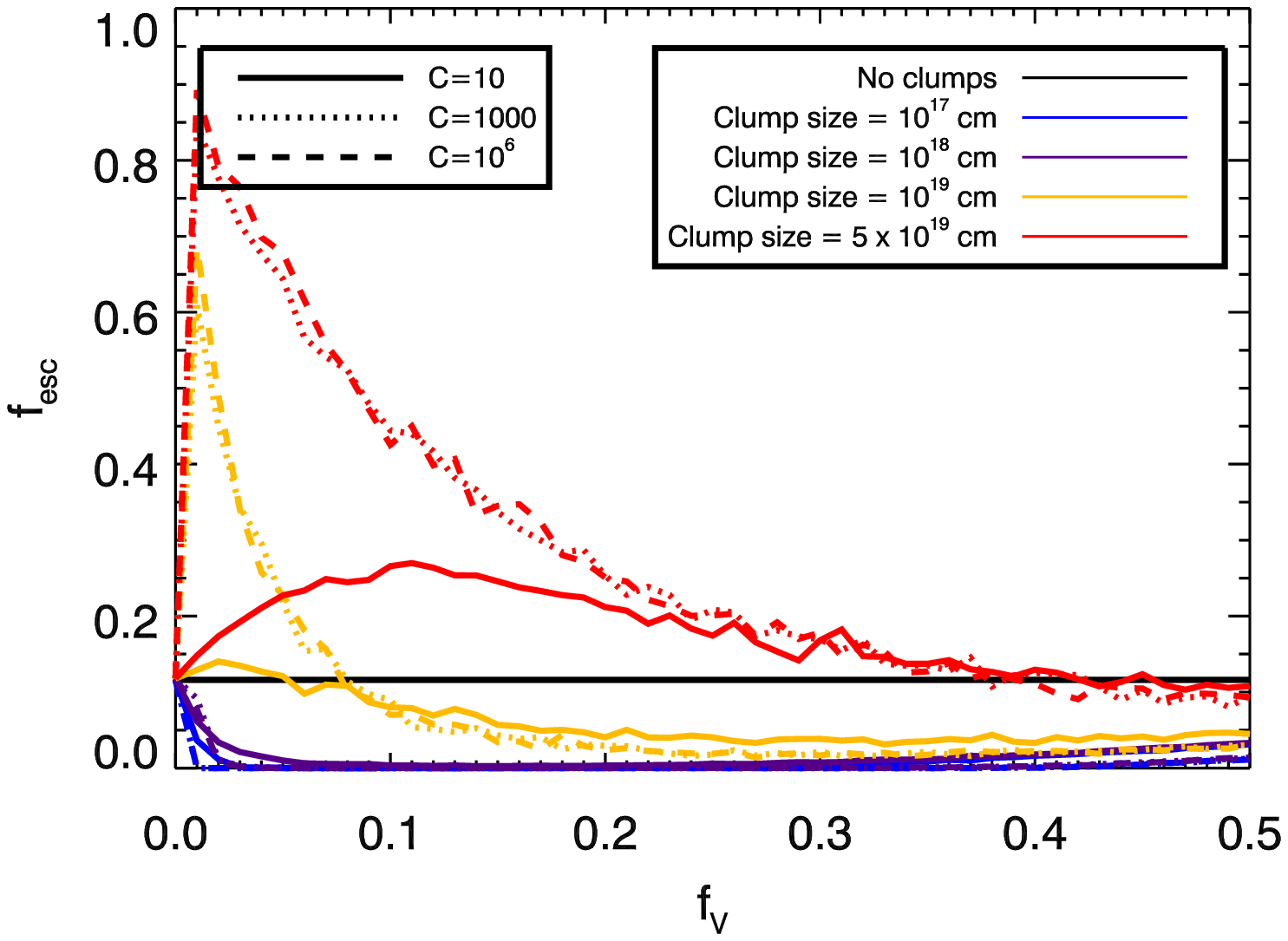}
\includegraphics[width=80mm]{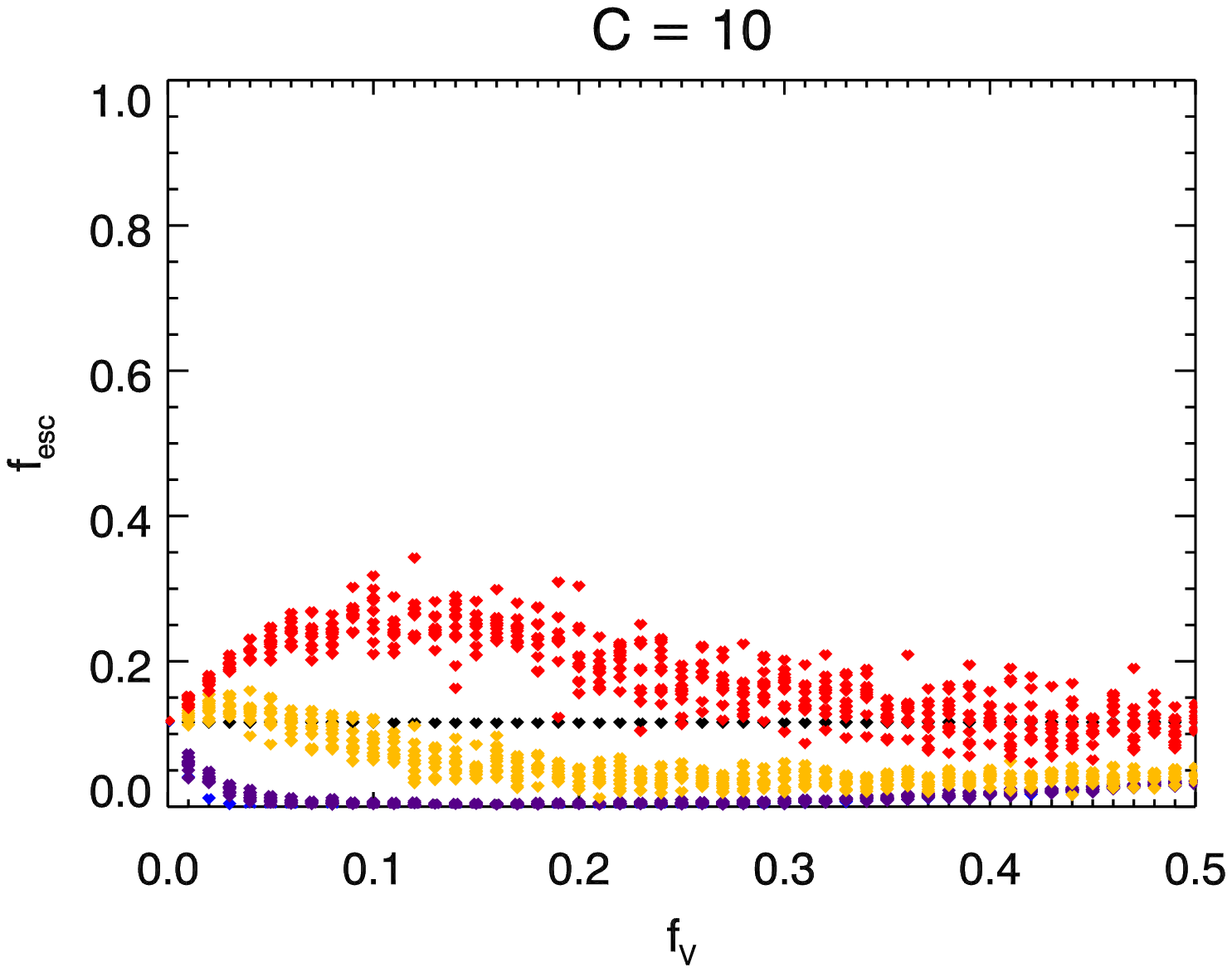}
\includegraphics[width=80mm]{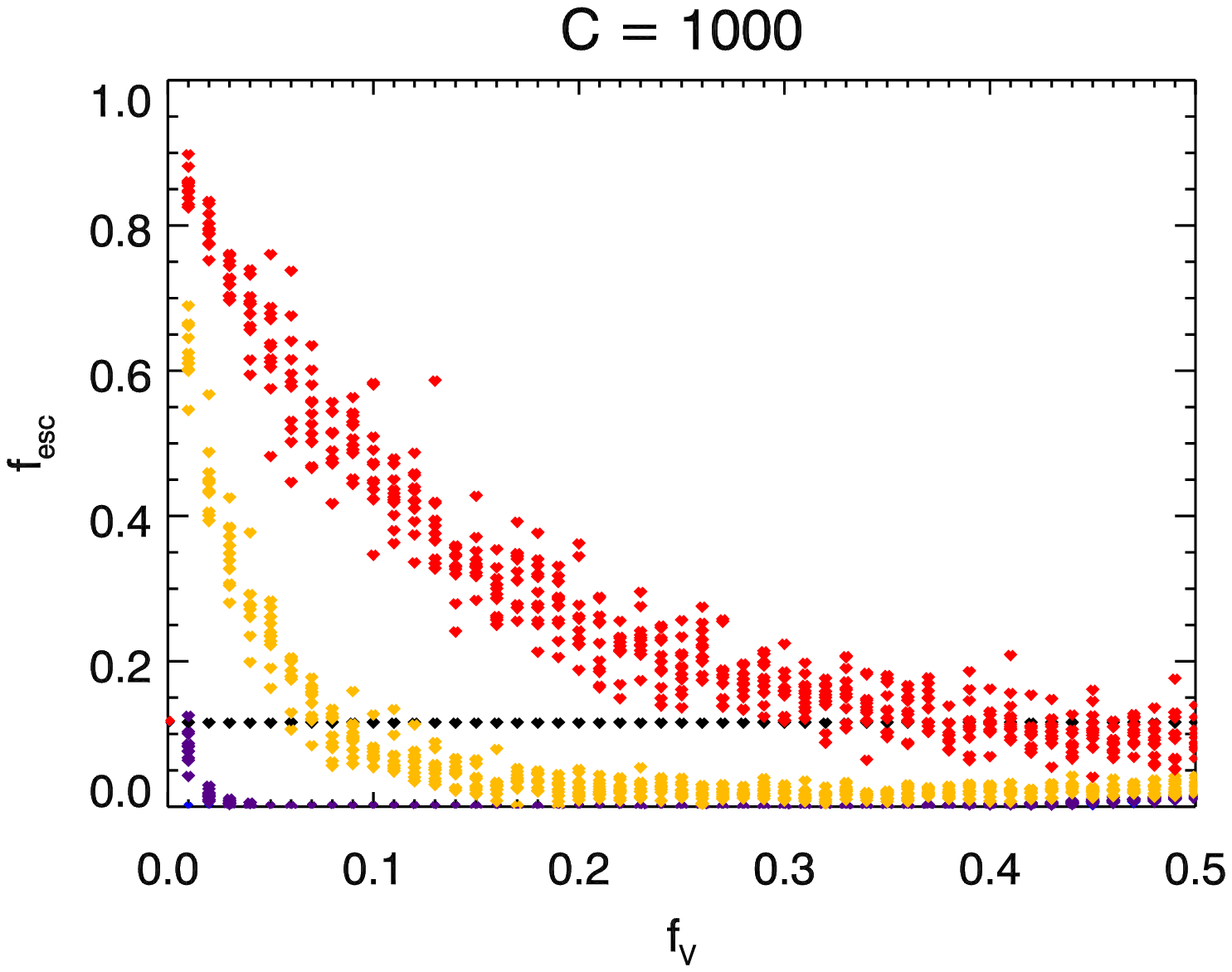}
\includegraphics[width=80mm]{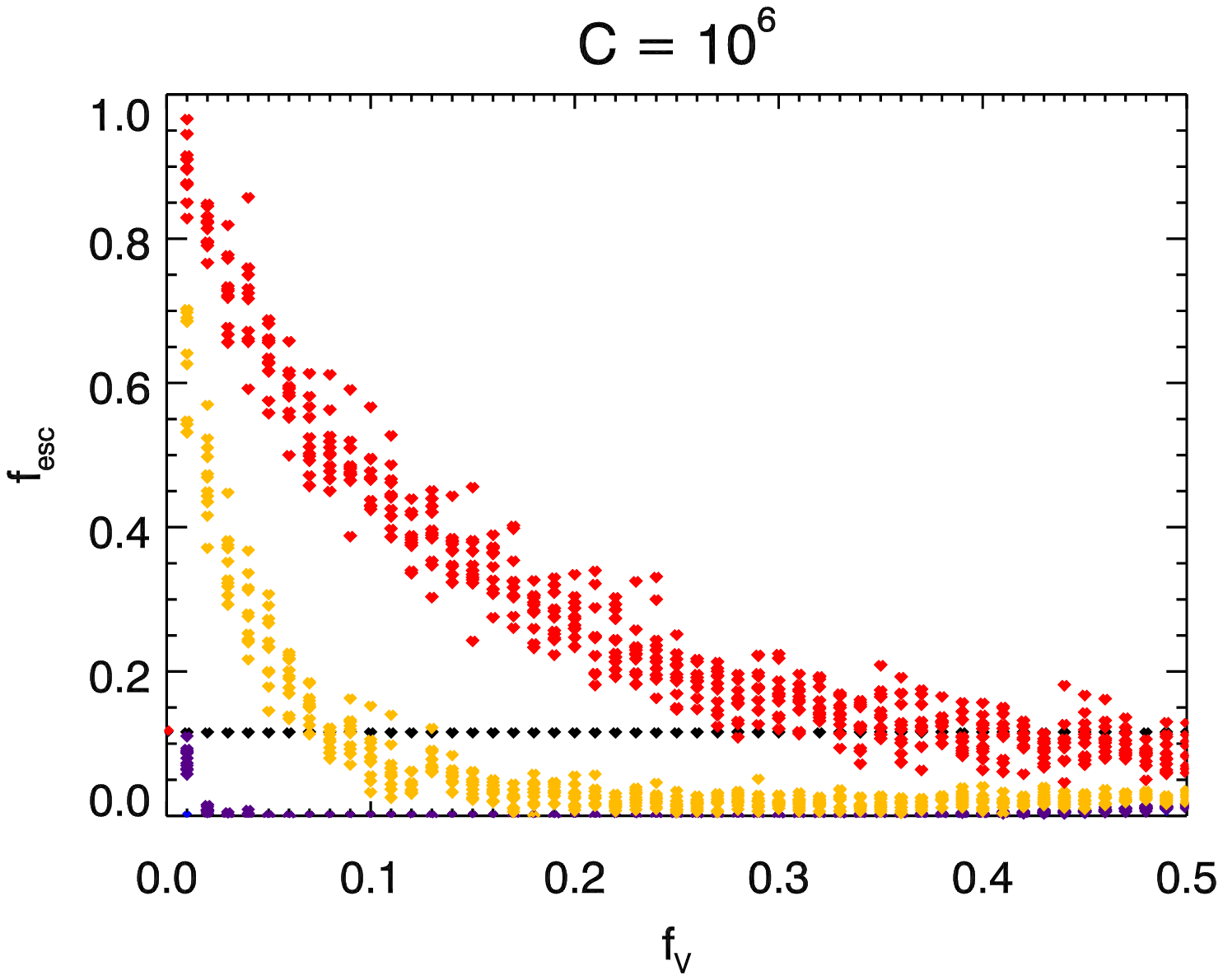}
\caption{
Effect of large clumps on escape fraction:   \fesc\ out of the disk is shown for a galaxy in a $10^9 M_\odot$ halo,
with $z_f = 10$, $f_* = 0.1$, and a Pop~III Larson initial mass spectrum, with $m_d=0.17$.  If the clumps are very
large and $f_V$ is very low, there are cases where \fesc\ is larger than the no-clump case.  In the top left corner, the results are averaged over ten runs to reduce noise.   In the remaining three panels, the distributions are shown for each run.  In each case, the black points represent a case with no clumps.  Galaxy properties are the same as in 
Figure \ref{fig:varyff}. 
 }
\label{fig:clumpsize}
\end{figure}

So far, we have only been exploring the results of small clumps
($10^{17}\: \rm{cm}$, or $\sim 0.3\: \rm{pc}$) in diameter.  What would happen if
we were to increase the size of these clumps?  In this case, a ray would
traverse fewer clumps as it travels out of the galaxy (the covering factor will fall), but any given clump
would be larger.  As shown in Figure \ref{fig:clumpsize}, $f_{esc}$ rises as the clumps
increase in size.  For very low values of $f_V$, only a few clumps exist
and not every ray comes in contact with a clump, increasing the escape
fraction above the case with no clumps.

\begin{figure}[t]
\centering \noindent
\includegraphics[width=95mm]{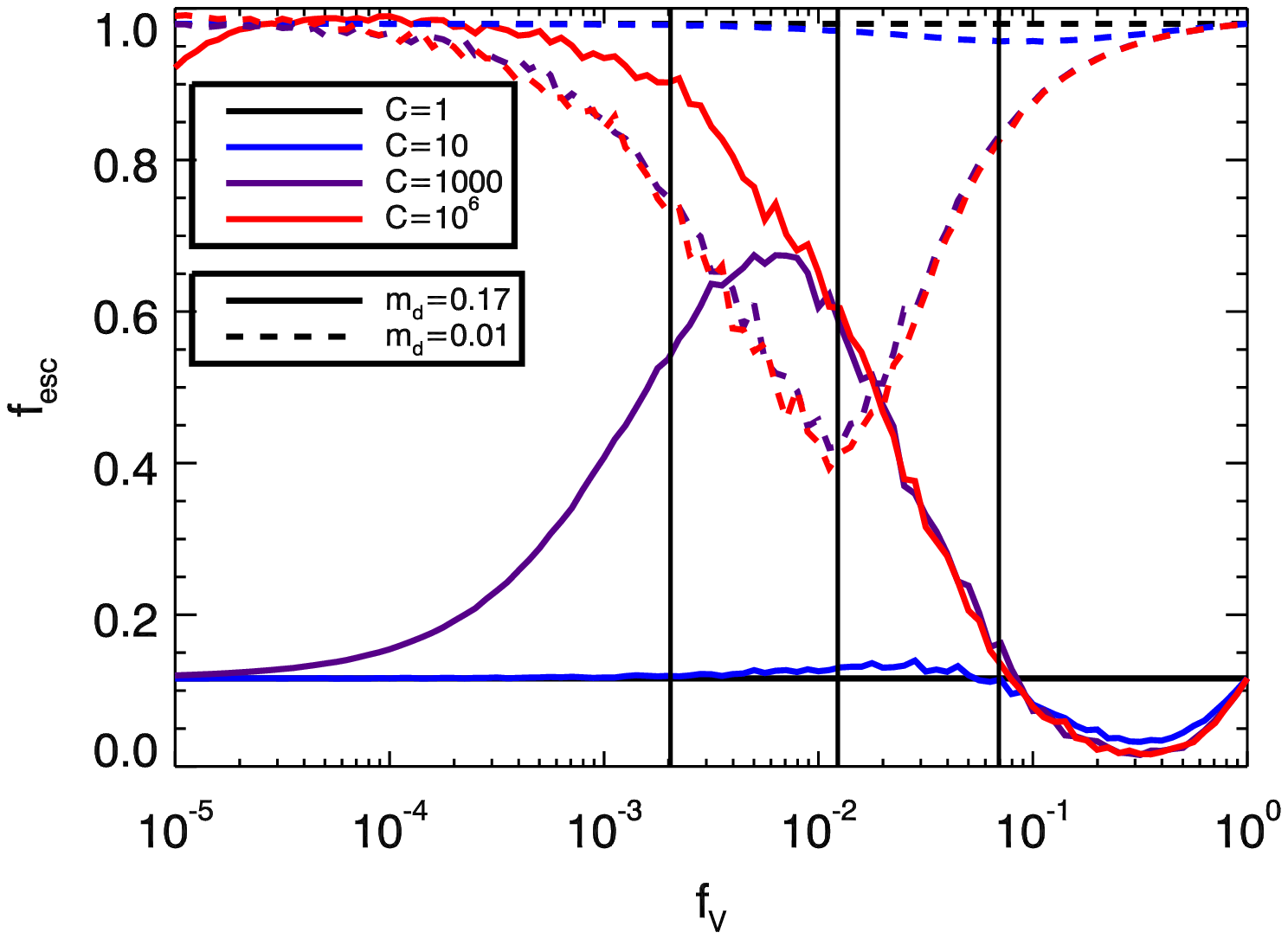}
\includegraphics[width=95mm]{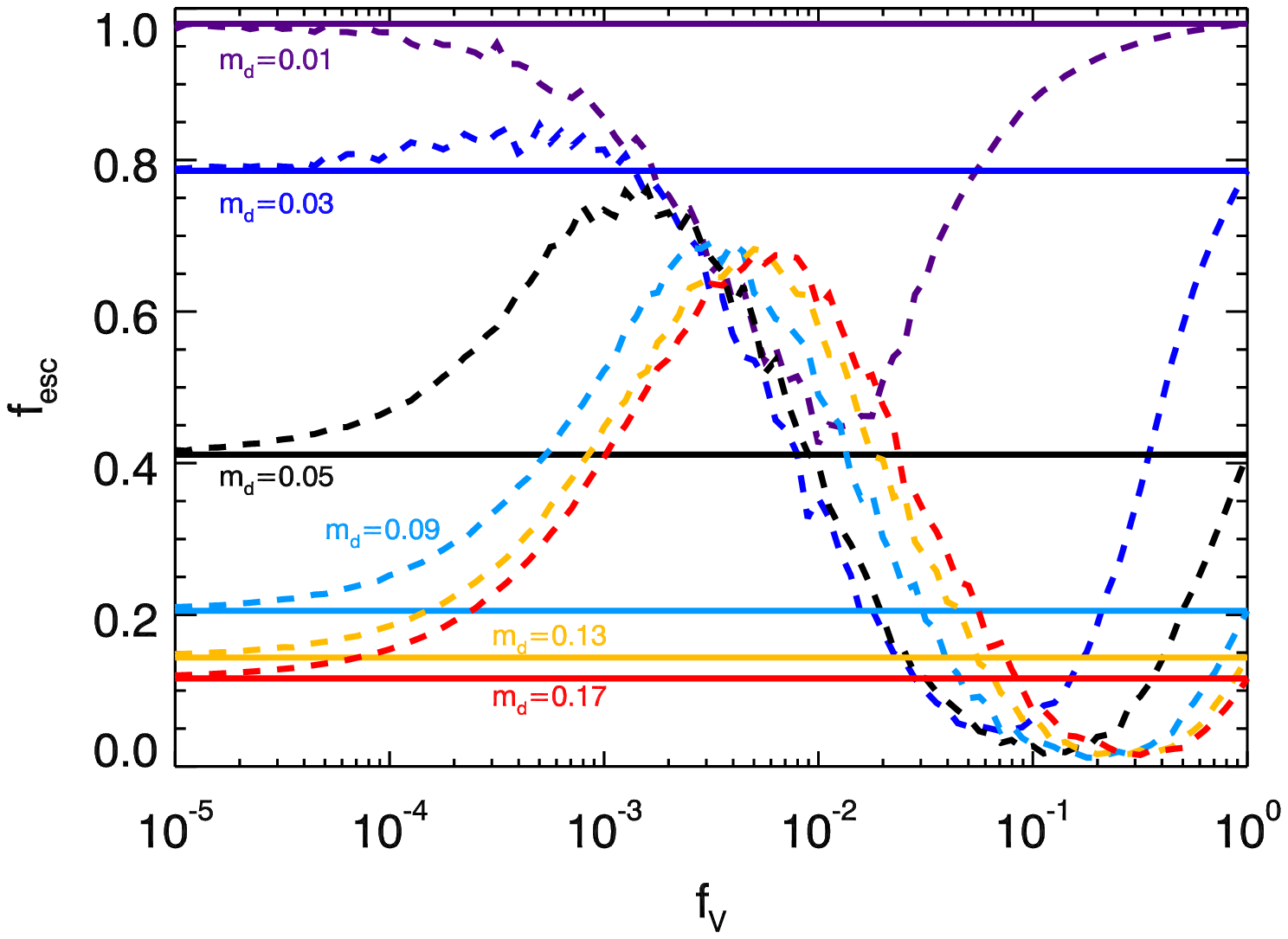}
\caption{%
{\it{Top:}} The \fesc\ out of the disk is shown for a galaxy in a $10^9 M_\odot$ halo,
with $z_f=10$, $f_* = 0.1$, and a Population III Larson initial mass spectrum.  The left-most solid vertical line represents the value of the volume filling factor $f_V$ needed for a photon transversing the longest path length to pass through an
average of one clump for a galaxy with $m_d = 0.17$, while the right-most solid vertical line represents the value of 
$f_V$ needed for a photon transversing the shortest path length to pass through an average of one clump for a galaxy with $m_d = 0.17$.  Therefore, to the right of this line, the covering factor is greater than one, and all path lengths intersect a clump, while to the left, there are clump-free path lengths out of the galaxy.  For a galaxy with $m_d=0.01$ the dashed vertical line represents the value of  $f_V$ needed for a photon transversing the shortest path length to pass through an average of one clump for a galaxy.  The value of $f_V$ needed for a photon traversing the longest path overlaps with the case where $m_d = 0.17$, because as $m_d$ changes, the scale height of the galaxy changes, but the radius remains the same.  {\it{Bottom}}: The dependence on the escape fraction as $m_d$ is varied.  The same population is shown, with $C = 1$ for the solid lines and $C = 1000$ for the dashed lines.  The results are averaged over ten runs to reduce noise.  
}
\label{fig:clumplog}
\end{figure}

\begin{figure}[t]
\centering \noindent
\includegraphics[width=120mm]{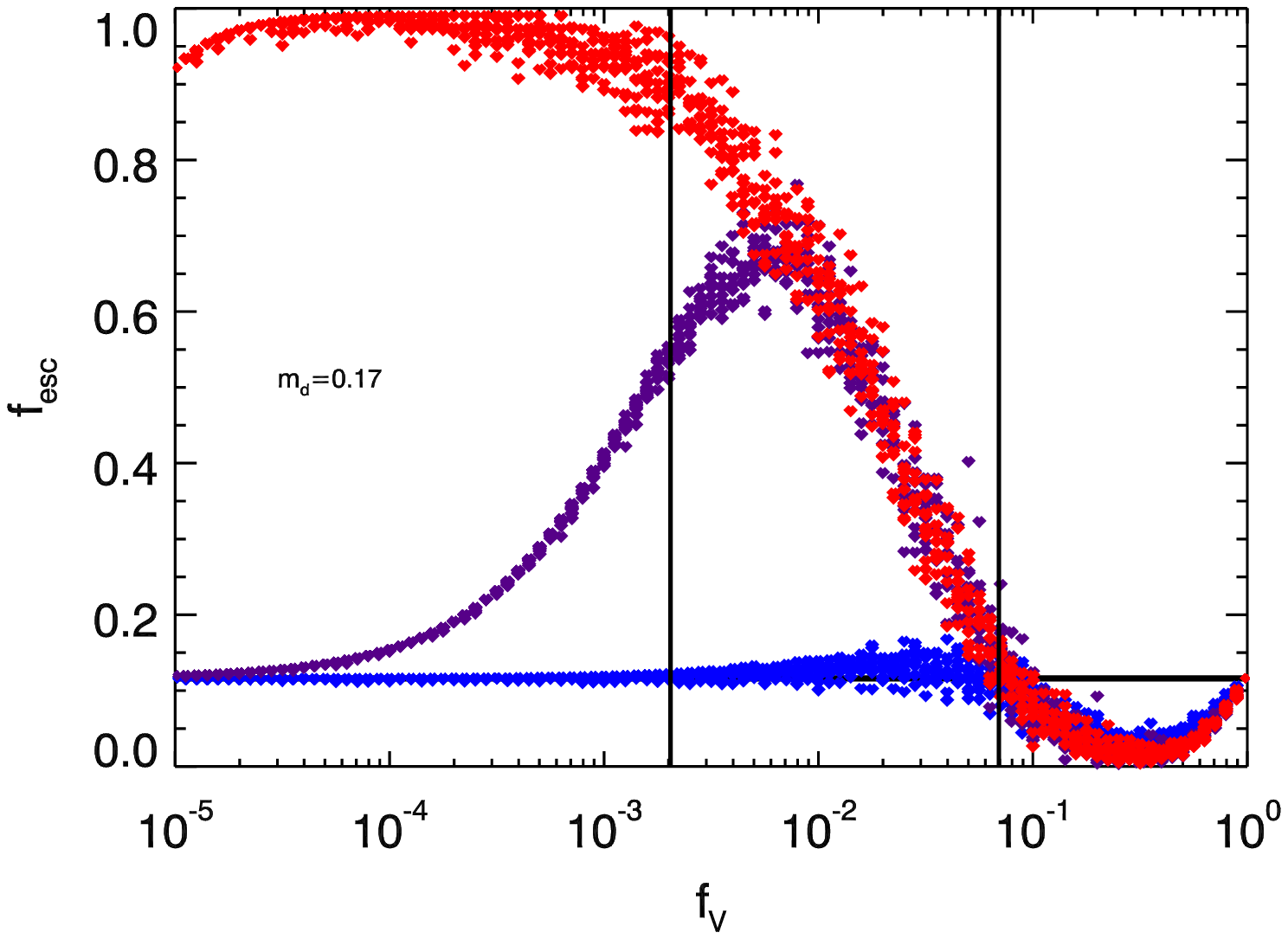}
\includegraphics[width=120mm]{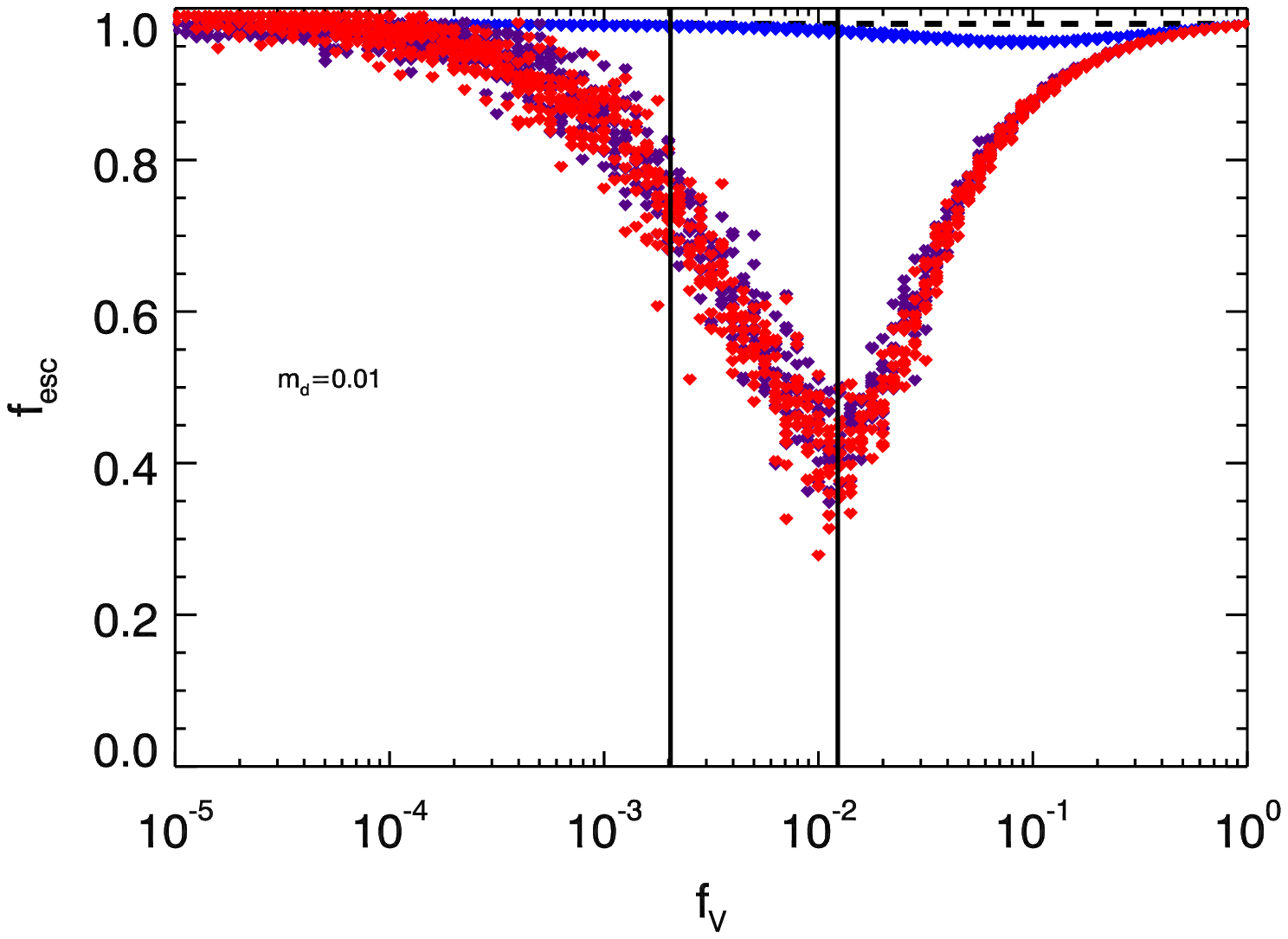}
\caption{Distribution of \fesc\ for each run shown in the top panel of Figure \ref{fig:clumplog}, with $m_d = 0.17$ in the top panel and  $m_d = 0.01$ in the bottom panel.  
}
\label{fig:clumplog2}
\end{figure}

To illustrate this further, Figure \ref{fig:clumplog} shows \fesc\ against $f_V$ for a large clump size ($10^{19}$~cm in diameter).  In the top panel, $C$ is varied for $m_d = 0.01$ and $m_d = 0.17$.  The left-most vertical line represents the value of $f_V$ needed for a photon traversing the longest path length to pass through an average of one clump for both $m_d=0.01$ and $m_d=0.17$. The right-most vertical solid line represents the value of $f_V$ for a photon traversing the shortest path length to pass through an average of one clump for $m_d=0.17$, while the dashed line is the shortest path for $m_d=0.01$.  To the right of these lines, all path lengths intersect a clump.  In other words, here the covering factor is greater than one. To the left of these lines, there are clump-free path lengths out of the galaxy.  For $m_d=0.17$, we see that if $f_V$ is low enough some rays will pass through fewer than one clump, on average, \fesc\ is much greater than the no-clump case.  For very low values of $f_V$, there are so few clumps that the interclump medium approaches the case with no clumps. Therefore, the plot of \fesc\ is peaked in the region where there are some paths that do not intersect a clump.  These results are averaged over ten runs, while Figure \ref{fig:clumplog2} shows the distribution for each run.

The bottom panel of Figure \ref{fig:clumplog} shows \fesc\ for various values of $m_d$.  As $m_d$ increases, the disk is less massive, and the escape fraction for a non-clumped galaxy rises.  For $m_d = 0.01$, the escape fraction of a non-clumped galaxy is $\sim 1$.  In this case, once clumps are added, the escape fraction decreases because regions dense enough to stop ionizing radiation are finally introduced.  When the covering factor is greater than one, the clumps grow less dense, causing the escape fraction to rise again.  For all other cases, \fesc\ peaks when the covering factor is less than one, and falls below the escape fraction for the non-clumped case when the covering factor is greater than one.


Star formation is more likely to take place in clumps.  Star formation will also have an effect on gas clumping
produced by stellar winds and supernova shells.  Because of this, clumps are likely to be distributed around locations of star formation.  To analyze this effect, we have defined a region 100~pc from the center of star formation in the galaxy.  Inside this region, the volume filling factor is $f_{\rm V,near}$, and outside of this region, the volume filling factor is 
$f_{\rm V,far}$.  For a case where clumps are only located near to the star formation center, $f_{\rm V,far} = 0$. 
If $f_{\rm V,far} < f_{\rm V,near}$, there are fewer clumps in the outer portions of the galaxy than near the star formation center.  

\begin{figure}[t]
\centering \noindent
\plotone{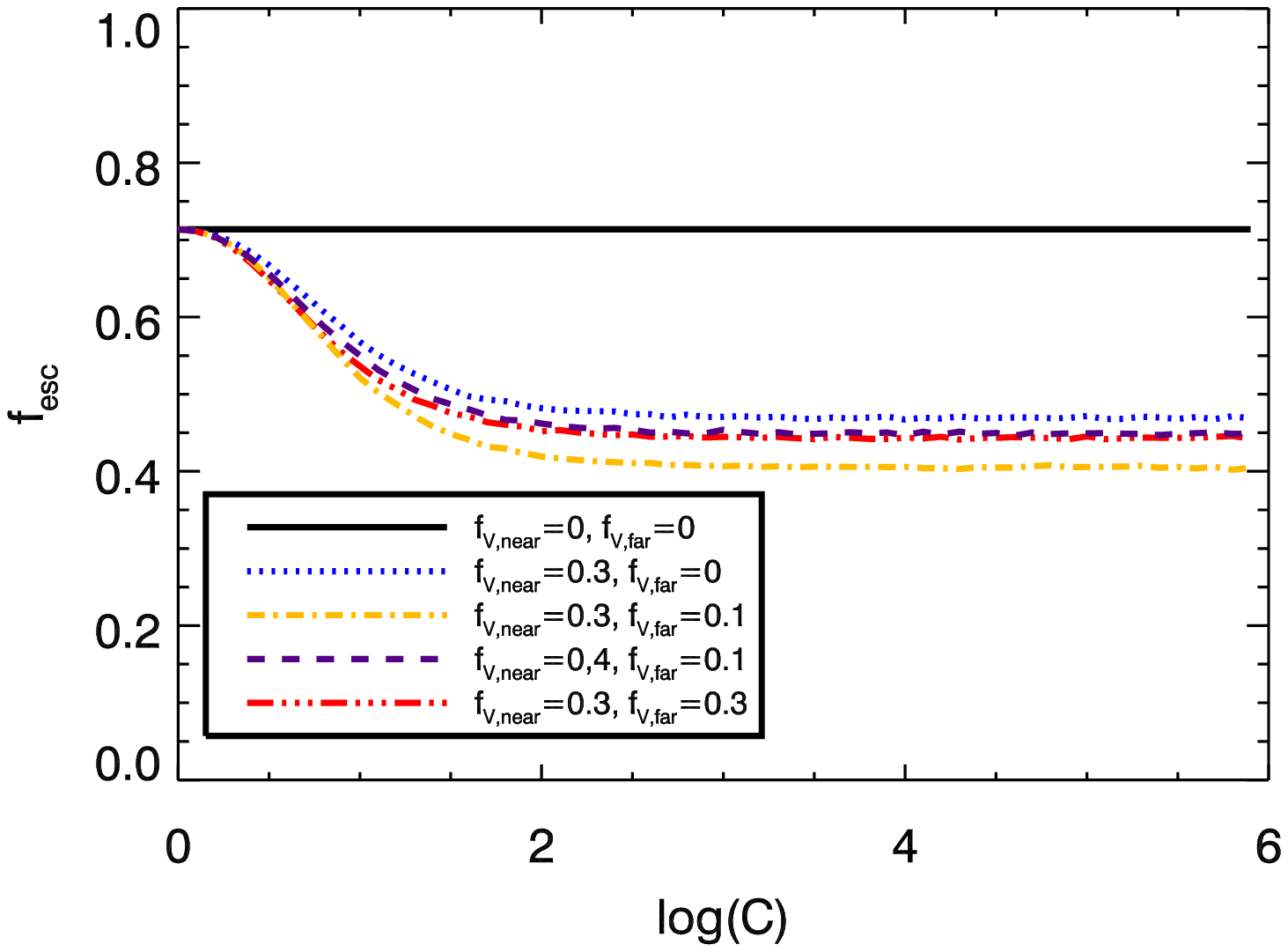}
\caption{%
Escape fraction when clumps are distributed inhomogeneously throughout the galaxy.   Two regions are defined, the region within 100~pc of star formation, which has a volume filling factor of $f_{\rm V,near}$, and the region outside of this, where the volume filling factor is $f_{\rm V,far}$.   Shown for a $10^9 M_\sun$ halo at $z_f=10$, with $f_* = 0.5$ and Pop~III stars with a Larson mass spectrum and $m_d = 0.17$.  Clumps are $10^{17}$~pc  across.  Please see the electronic edition for a color version of this plot.
}
\label{fig:clumpdist}
\end{figure}

Results are shown in Figure \ref{fig:clumpdist}.  The black solid line shows the case with no clumps, where $f_{V}=0$ throughout, and the red triple-dot dashed line represents the case where $f_V=0.3$ throughout.  For the case where 
$f_{\rm V,near}=0.3$ and $f_{\rm V,far}=0$, clumps are only near stars.  This results in a higher escape fraction than if the entire galaxy had $f_{V}=0.3$.  When there are some clumps far from star formation, in the case where $f_{\rm V,near}=0.3$ and $f_{\rm V,far}=0.1$, the escape fraction falls below the case where if the entire galaxy had $f_{V}=0.3$.  This is a result of clumps becoming more dense as $f_V$ increases.  

We have chosen a 100~pc  radius of the region near to the star formation.  If this region is much larger, $\sim 1$~kpc, the escape fraction will be equal to the case $f_V = f_{\rm V,near}$, where $f_V$ is the volume filling factor if the galaxy has the same value throughout.  If the region is much smaller, $\sim10$~pc, the escape fraction will be equal to the case where $f_V = f_{\rm V,far}$.  Overall, the distribution of clumps does not change the escape fraction significantly.


\subsection{Properties of Stars and the Galaxy}
\begin{figure}[t]
\centering \noindent
\includegraphics[width=120mm]{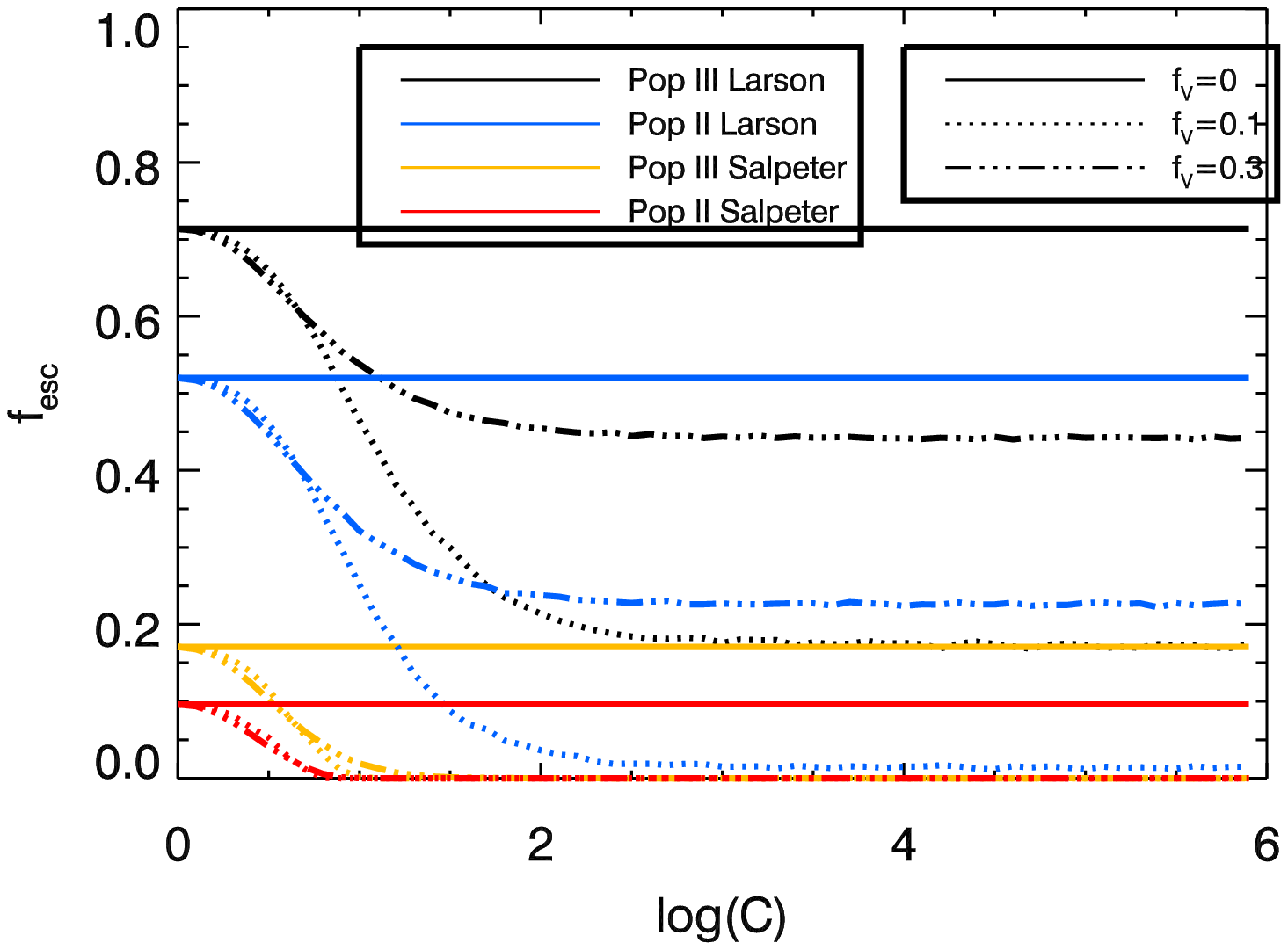}
\includegraphics[width=120mm]{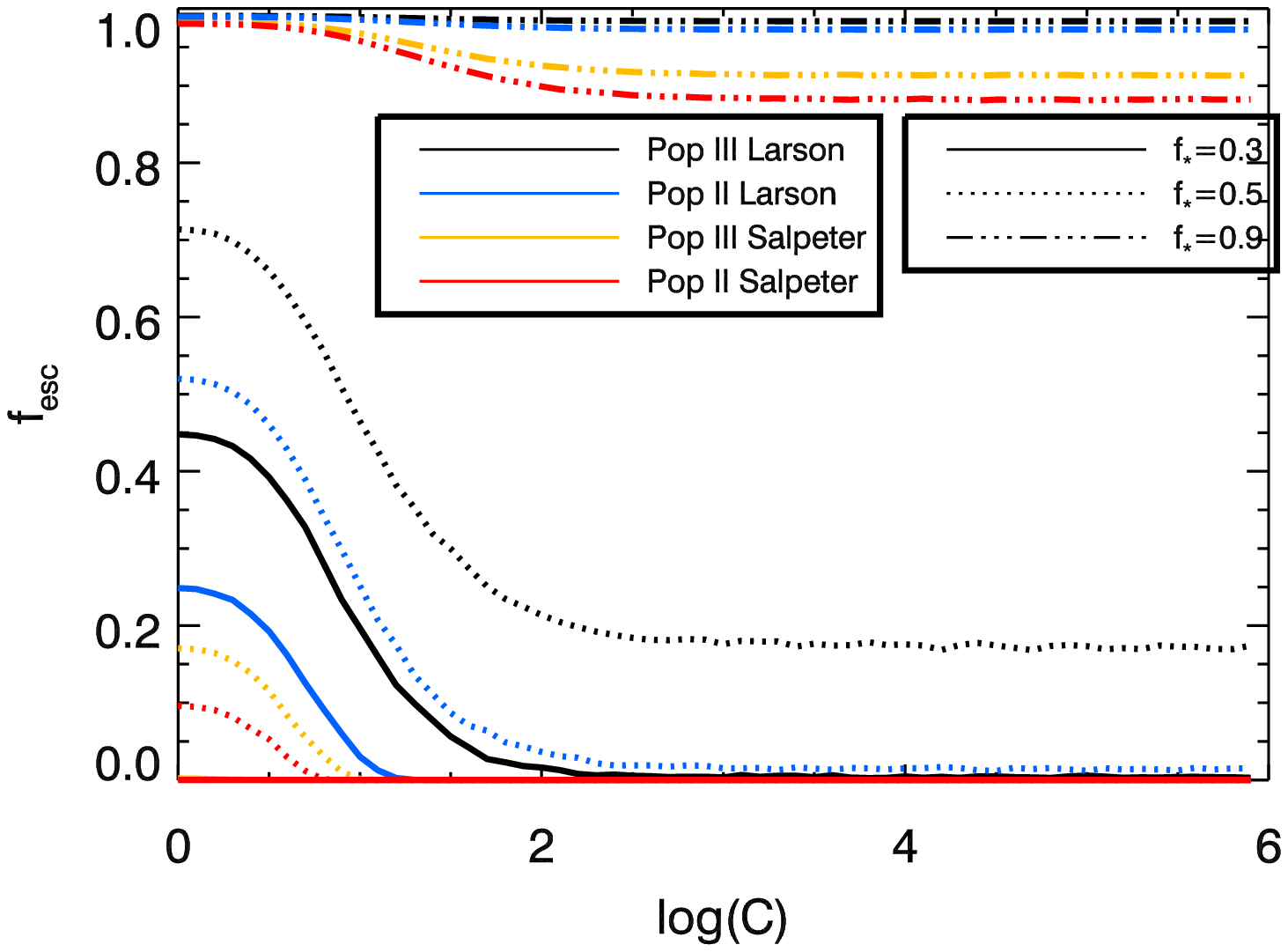}
\caption{%
The $f_{esc}$ for the disk is shown for stars of varied masses and metallicities and various values of $f_V$ with 
$f_* = 0.5$ ({\it{top panel}}) and various values of $f_*$  for $f_V=0.1$ ({\it{bottom panel}}).  Very high values of 
$f_*$ ($0.9$) approach a case in which all ionizing photons are escaping.  Both are shown for a $10^9 M_\sun$ halo at $z_f=10$.  In each case, the highest line is for Population III Larson, followed by Population II Larson, Population III Salpeter, and Population II Salpeter.  For $f_* =0.3$, \fesc\ $\sim 0$ for both Salpeter cases.  Please see the electronic edition for a color version of this plot.
}
\label{fig:Q1}
\end{figure}

In Figure \ref{fig:Q1}, we analyze how the stellar population affects the escape fraction.  In the top panel, $f_*$ is held constant as $f_V$ increases.  In the bottom panel, $f_V$ is held constant as $f_*$ increases. Both plots show metal-free (Population III) stars and metal-poor (Population II) stars, as well as stars with a heavy Larson initial mass spectrum and a light Salpeter initial mass spectrum.  In both cases, \fesc\ is proportional to the number of ionizing photons that are emitted by the stars, with heavier stars or stars with fewer metals more likely to produce photons that can escape the nebula.  This is because when more ionizing photons are produced, the critical angle where photons can break free from the halo increases, and hence more photons escape.  When the galaxy forms more stars (higher $f_*$), there is the added effect of less hydrogen remaining in the galaxy to absorb ionizing photons.   Therefore, \fesc\ increases greatly as $f_*$ increases.


\begin{figure}[t]
\centering \noindent
\plotone{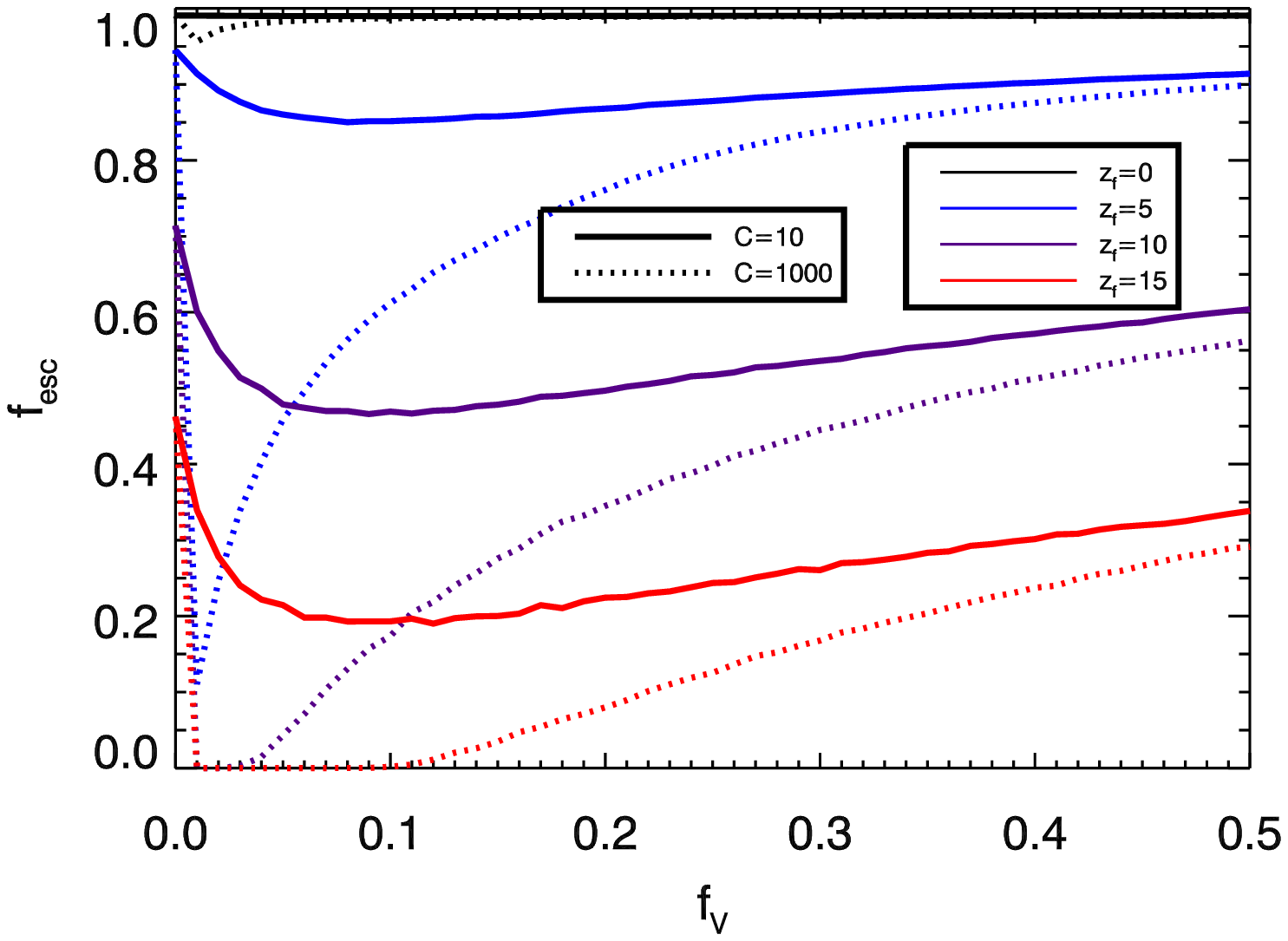}
\caption{%
The \fesc\ for the disk is shown for various values of $z_f$.  Shown for a halo with Pop~III stars with a Larson 
mass spectrum and $f_* =0.5$.  Please see the electronic edition for a color version of this plot.
}
\label{fig:galaxy1}
\end{figure}

The star formation redshift, $z_f$, is varied in Figure \ref{fig:galaxy1}.  
As $z_f$ increases, the galaxy is smaller and more concentrated.  Therefore, it is easiest for
photons to escape from less dense disks at low redshifts.  At redshifts where we expect reionization to take place, it is harder for photons to escape the galaxy.  This problem may be remedied by the high-redshift expectation of
more massive or metal-free stars with higher values of $f_*$.


\section{COMPARISON TO PREVIOUS LITERATURE}
\label{sec:Lit}

As noted in the introduction, there have been many previous studies that calculated the number of ionizing photons
emitted from high redshift halos, resulting in a wide range of values for the escape fraction. Various factors are proposed that affect the number of ionizing photons that escape into the IGM, in particular the effects of a clumpy ISM.  
\citet{boisse:1990} and \citet{witt/gordon:1996} found that clumps increase transmission, and
\citet{hobson/scheuer:1993} found that a three-phase medium (clumps grouped together, rather than randomly distributed) further increases transmission.  Very dense clumps (with $C=10^6$) were studied by
\citet{wood/loeb:2000}, who found that clumps increase \fesc\ over the case with no clumps.  For very small values of
$f_V$, their \fesc\ was very high, because most of the density is in a few very dense clumps, and most lines of sight do not encounter a clump.  Their clump size is 13.2~pc, which is similar to our largest clump size, 
$5 \times 10^{19}$~cm.  Their results are consistent with our findings for clumps with large radius, low $f_V$,
and high $C$, where most rays do not encounter a clump. 

\citet{ciardi/etal:2002} included the effect of clumps using a fractal distribution of the ISM with $C = 4-8$.  They noted that this distribution of clumps increases \fesc\ in cases with lower ionization rate because there are clearer sight lines.  They found that \fesc\ is more sensitive to the gas distribution than to the stellar distribution.  

\citet{dove/shull/ferrara:2000} reported that \fesc\ decreases as clumps are added.  This results from the fact that
adding clumps does not change the density of the interclump medium.  In their model, as clumps are added, the mass of hydrogen in the galaxy increases.  On the other hand, our current method decreases the density of the interclump medium as clumps are added or become denser to keep the overall mass of the galaxy constant. Photons
are more likely to escape along paths with lower density, as in irregular galaxies and along certain lines of sight
\citep{gnedin/etal:2008,wise/cen:2008}.   Shells, such as those created by supernova remnants (SNRs), can trap ionizing photons until the bubble blows out of the disk, allowing photons a clear path to escape and causing \fesc\
to rise \citep{dove/shull:1994,dove/shull/ferrara:2000,fujita/etal:2003}.  These SNRs or superbubbles create porosity in the ISM, and above a critical star formation rate, \fesc\ rises \citep{clark/oey:2002}.  This is similar to what is seen in our results.  As with a dense clump, a shell will essentially stop all radiation, while a clear path, similar to the case with a low $f_V$, allows many free paths along which radiation can escape.  Our model extends the previous work by enlarging the parameter space.  We see how low values of $f_V$, changing the clump size, and location of the clumps affect the results.  This heavy dependence on how the clumping can affect the escape fraction can explain why such a variation is seen in observations in the escape fraction from galaxy to galaxy.

Previous work differs as to whether or not \fesc\ increases or decreases with redshift.  
\citet{ricotti/shull:2000} state that \fesc\ decreases with increasing redshift for a fixed halo mass,
but is consistent with higher escape fraction from dwarf galaxies.  This assumes that the star formation efficiency is proportional to the baryonic content in a galaxy.  However, other studies seem to indicate that
\fesc\ increases with redshift.   High resolution simulations of \citet{razoumov/sl:2006} state that \fesc\
increases with redshift from $z = 2.39$ (where \fesc\ = 0.01--0.02) to $z = 3.8$ (where \fesc\ = 0.06--0.1).
This is a result of higher gas clumping and lower star formation rates at lower redshifts, causing the escape fraction to fall.  At higher redshifts, the simulations of \citet{razoumov/sl:2009} see this trend continue, with an \fesc\ $\approx 0.8$
at $z \approx 10.4$ that declines with time.  This trend has also been seen observationally from $0<z<7$  \citep{bridge:2010,cowie/etal:2009,inoue/etal:2006,Iwata/etal:2009, Siana/etal:2007, Bouwens/etal:2010}.  (However, \citet{vanzella/etal:2010} point out that observational measurements of the escape fraction can be contaminated by lower redshift interlopers.)  
Other simulations have given different results.  \citet{gnedin/etal:2008} say that \fesc\ changes
little from $3 < z < 9$, always being about 0.01--0.03.  This difference could possibly be attributed to how the models deal with star formation efficiencies within galaxies.  \citet{wood/loeb:2000} state that since disk density increases with redshift, \fesc\ will fall as the formation redshift increases, ranging from \fesc\ = 0.01--1.  We found that \fesc\
decreases with increasing redshift of formation, since disks are more dense.  However, this assumes that the types of stars and $f_*$ remain constant.  If $f_*$ is larger at higher redshifts and if stars are more massive and have a lower metallicity (likely), the number of ionizing photons should increase, which could cause \fesc\ to increase, despite a denser disk.  It is also possible that the disk morphology does not exist at high redshifts.  Therefore, in order to understand the evolution of \fesc\ with redshift, one must understand the evolution of other properties, namely, $f_*$, $Q_{\rm pop}$, and the density distribution of gas within a galaxy.


\section{CONSTRAINTS FROM REIONIZATION}
\label{sec:reion}

If galaxies are responsible for keeping the universe reionized,  there must be a minimum number of photons that can escape these galaxies to be consistent with reionization.  The star formation rate ($\dot{\rho}$) that corresponds to a star formation efficiency $f_*$ is given by:
\begin{equation}
\dot{\rho}(z) = 
    0.536~M_\sun~{\rm yr^{-1}}~{\rm Mpc}^{-3}
      \left(\frac{f_*}{0.1}\frac{\Omega_bh^2}{0.02}\right)
    \left(\frac{\Omega_mh^2}{0.14}\right)^{1/2} \left(\frac{1+z}{10}\right)^{3/2}
       y_{\rm min}(z) e^{-y_{\rm min}^2(z)/2}
\label{eq:ferKom}
\end{equation}
\citep{fernandez/komatsu:2006}, assuming a Press-Schechter mass function
\citep{press/schechter:1974}, with 
\begin{equation}
     y_{\rm min}(z) \equiv \frac{\delta_c}{\sigma[M_{\rm min}(z)]D(z)}  \; .
\end{equation}
Here, $\delta_c$ is the overdensity, $\sigma(M)$ is the present-day rms amplitude of mass
fluctuations, and $M_{\rm min}$ is the minimum mass of halos that create stars.

\noindent Similarly, the \fesc\ needed to reionize the universe can be related to the critical star formation rate, 
$\dot{\rho}_{\rm crit}$, needed to keep the universe ionized:
\begin{equation}
\dot{\rho}_{\rm crit}(z) = (0.012 \: M_\sun \: {\rm{yr}}^{-1} \: {\rm{Mpc}}^{-3} \: ) 
   \left[\frac{1+z}{8}\right]^3\left[\frac{C_H/5}{f_{esc}/0.5}\right]\left[\frac{0.004}{Q_{\rm LyC}}\right]T_4^{-0.845}
\label{eq:trenti}
\end{equation}
\citep{madau/etal:1999,trenti/shull:inprep}, 
which results from the number of photoionizations needed to balance recombinations
to keep the universe ionized at $z=7$. Here, $C_H$ is the clumping of the IGM (which we scale to a typical value
of $C_H=5$), $T_4$ is the temperature of the IGM in units of $10^4$~K, and $Q_{\rm LyC}$ is the conversion factor from $\dot{\rho}(z)$ to the total number of Lyman continuum photons produced per $M_{\odot}$ of star formation,
\begin{equation}
Q_{\rm LyC} \equiv \frac{N_{\rm LyC}/10^{63}}{\dot{\rho}_{\rm crit}t_{\rm rec}},
\end{equation}
where $N_{\rm LyC}$ is the number of Lyman continuum (LyC) photons produced by a star and $t_{\rm rec}$ is the hydrogen recombination timescale.  We assume that $Q_{\rm LyC}=0.004$, which is reasonable for a low-metallicity stellar population.

\begin{figure}[t]
\centering \noindent
\includegraphics[width=120mm]{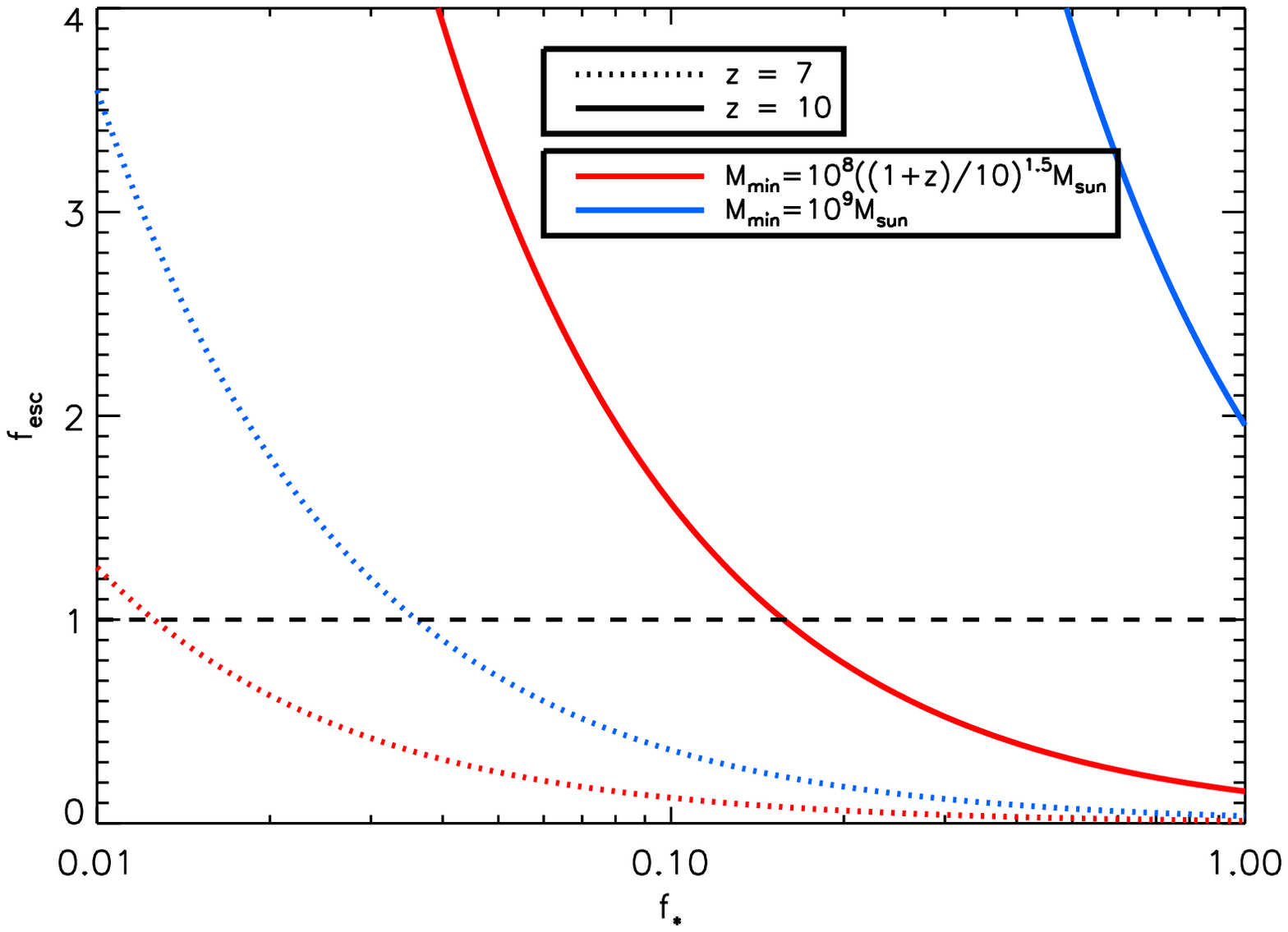}
\includegraphics[width=120mm]{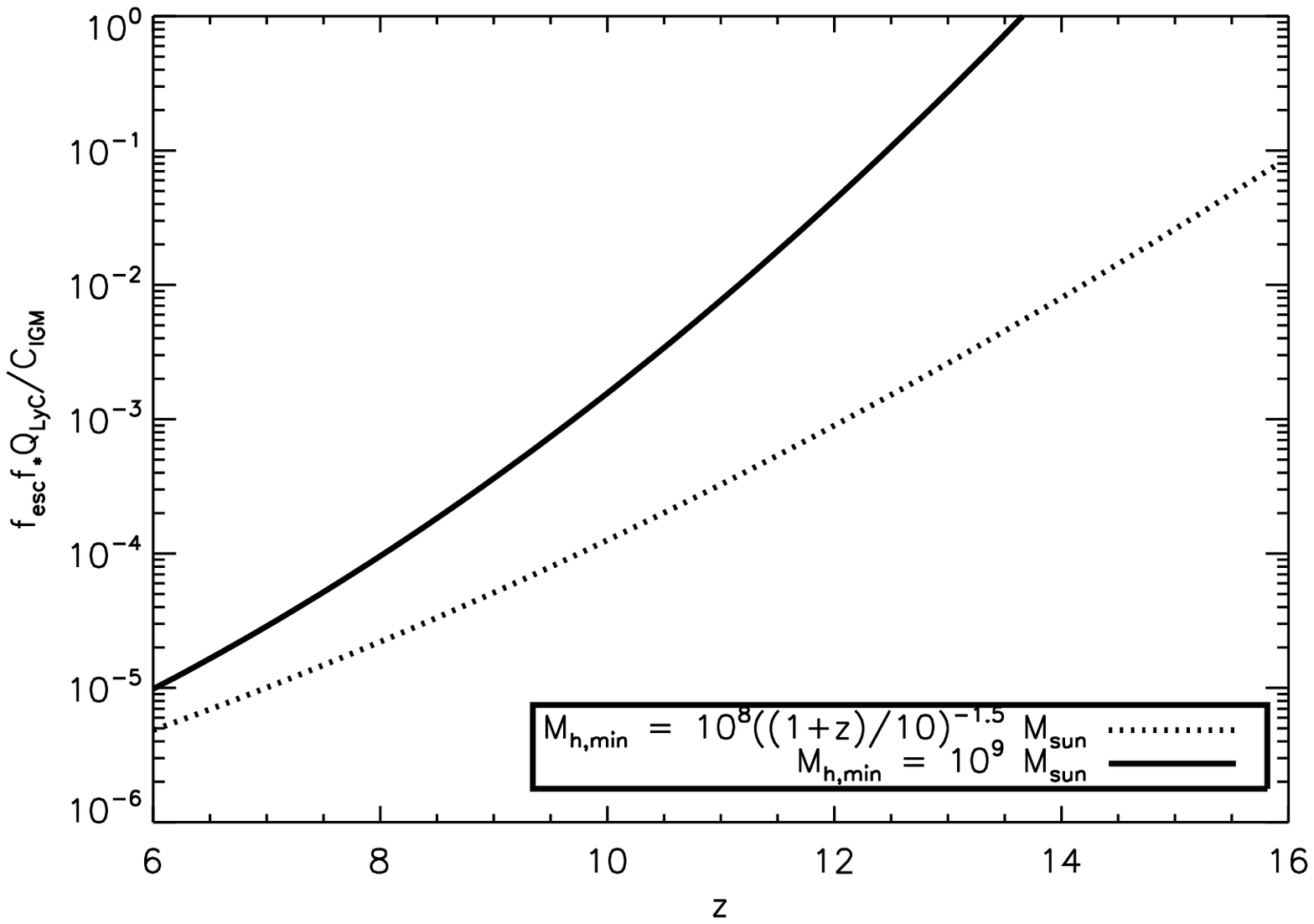}
\caption{%
{\it{Top:}}  The \fesc\ needed from a population of galaxies with various values of
$f_*$ to reionize the universe at $z = 7$ or $10$.  If the required \fesc\  lies above $1$
(dashed line), the population cannot reionize the universe.  {\it Bottom:}  The product 
$f_{\rm esc} f_*Q_{LyC}/C_{IGM}$, which is directly constrained by reionization.  The 
region above each line has a value of $f_{\rm esc}f_*Q_{LyC}/C_{IGM}$ large enough
 to reionize the universe.  Please see the electronic edition for a color version of this plot. 
}
\label{fig:reion1}
\end{figure}

By requiring the star formation rate to be at least as large as the critical value, we can solve for the value of
\fesc\ needed to reionize the universe.  Results are shown in the top panel of Figure \ref{fig:reion1} plotted at 
two redshifts, $z=7$ and $z=10$.  We also consider stars forming in galaxies down to a minimum mass
\begin{equation}
M_{min} = (10^8~M_{\odot})  \left[ \frac{(1+z)}{10}\right] ^{-1.5}  \; , 
\end{equation}
\citep{barkana/loeb}, or if smaller halos are suppressed and only those above $10^9M_\sun$ are forming stars.  If the required \fesc\ exceeds $1$ (shown by the dashed line), the given population cannot reionize the universe.  As redshift decreases, it becomes much easier to keep the universe reionized, and a smaller \fesc\ is needed, as expected.
 
At higher redshifts, it is harder for stars to keep the IGM ionized because the gas density is higher and there are fewer massive halos forming stars.  If small halos are suppressed, the remaining high-mass halos have a much harder time keeping the universe ionized.  It is interesting to note that the universe cannot be reionized at $z = 8-10$ if only
larger halos ($M_h > 10^9 M_\sun$) are producing ionizing photons that escape into the IGM.  At $z = 9$, $f_*$ must always be above $0.1$ for all cases shown in order for the universe to be reionized.  At $z=7$, $f_*$ can be very low. 
Alternately, one can set equations \ref{eq:ferKom} and \ref{eq:trenti} equal and constrain directly the value of $f_{esc}f_*Q_{LyC}/C_{IGM}$.  Any values greater than this would be able to reionize the universe.  This is shown in the bottom panel of Figure \ref{fig:reion1}.

Observations of high redshift galaxies at $z = 7-10$ \citep{robertson/etal:2010,Bouwens/etal:2010}
show that currently observed galaxies, along with an escape fraction that is compatible with lower redshift observations (\fesc\ $\approx 0.2$), are not able to reionize the universe.  Therefore, the faint end slope of the luminosity function must be very steep to create a greater contribution to reionization from high-redshift small galaxies, the escape fraction at high redshifts was probably much higher, the IMF was top heavy and made of low metallicity stars, and/or the clumping factor of the IGM was low  \citep{bunker/etal:2004,bunker/etal:2010, gnedin:2008, gonzalez/etal:2010, lehnert/bremer:2003,mclure/etal:2010, oesch/etal:2010a, oesch/etal:2010b, ouchi/etal:2009, richard/etal:2008, stiavelli/etal:2004, yan/windhorst:2004}.  In addition, observations of Ly$\alpha$ absorption toward quasars and the UV luminosity function suggest that \fesc\ is allowed to be much lower if low-mass galaxies are allowed to be sources of ionizing photons \citep{srbinovsky/wyithe:2008}.   This is consistent with our findings;  small galaxies need to be included and their star formation not suppressed to permit \fesc\  to be sufficiently high to ionize the universe.  Low values of \fesc, consistent with the low redshift universe, would require high, and in some cases unreasonably high, values of $f_*$.   Therefore, the escape fraction was almost certainly higher in the past than it is today. 


\section{CONCLUSIONS}
\label{sec:Conclusions}

We have explored how the internal properties of galaxies can affect the amount of escaping ionizing radiation.  The properties of clumps within the galaxy have the strongest effect on \fesc.     
When the covering factor is much less than one, only  a few, dense clumps exist.  These clumps only stop a small fraction of the ionizing radiation, and therefore, the escape fraction will be large.  As the covering factor increases to one, the escape fraction will  fall, eventually to become smaller than the case with a non-clumped galaxy.  (The exception to this is if the non-clumped galaxy has an escape fraction of about one.  In this case the addition of clumps will only cause the escape fraction to fall around a covering factor of one.)  This indicates that the escape fraction is very sensitive to the way the ISM is distributed, causing the escape fraction to vary from 1 to 0 in some cases.  

The escape fraction depends on the star formation efficiency and the population of stars formed at the center of the galaxy. As the number of ionizing photons increases, the critical angle at which photons can escape the galaxy will also increase, which will have a direct effect on the escape fraction.  Since disks were more likely more dense at higher redshifts, the escape fractions will be lower, unless a change in the star formation efficiency or stellar population would increase the number of ionizing photons.  Therefore, the escape fraction depends directly on the number of ionizing photons and the distribution of hydrogen.  These variations can be extreme and can help to explain why some observations see large differences in the escape fraction from galaxy to galaxy.  Other factors, such as the mass of the galaxy and the redshift, affect the escape fraction indirectly (as the number of ionizing photons and distribution of hydrogen will change with mass or redshift).

At high redshifts, small galaxies are probably needed to help complete reionization, consistent with recent observations.  In addition, low values of the escape fraction, $f_{\rm esc} \approx 0.02$, similar to what is seen at low redshifts, are probably insufficient to allow reionization to be completed.  Otherwise, they would require very high values of the star formation efficiency.

\acknowledgements

We acknowledge support from the University of Colorado Astrophysical Theory
Program through grants from NASA (NNX07AG77G) and NSF (AST07-07474).


\newpage

\begin{thebibliography}{99}

\bibitem[Barkana \& Loeb(2001)]{barkana/loeb}
Barkana, R., \& Loeb, A. 2001,
Phys. Rep., 349, 125
\bibitem[Bergvall et al.(2006)]{bergvall/etal:2006}
Bergvall, N.,  Zackrisson, E., Andersson, B.-G., Arnberg, D., Masegosa, J., \& …stlin, G.  2006,
A\&A, 448, 513
\bibitem[Bland-Hawthorn \& Maloney(2002)]{bland/maloney:2002}
Bland-Hawthorn, J., \& Maloney, P. R.  2002,
ASPC, 254, 267
\bibitem[Bland-Hawthorn \&  Maloney(1999)]{bland/maloney:1999}
Bland-Hawthorn, J., \& Maloney, P. R., 1999,
ApJ, 510, L33
\bibitem[Boiss\'{e}(1990)]{boisse:1990}
Boiss\'{e}, P. 1990,
A\&A, 228, 483
\bibitem[Bouwens et al.(2010)]{Bouwens/etal:2010}
Bouwens, R. J., et al. 2010,  
\apj, 708, L69
\bibitem[Bridge et al.(2010)]{bridge:2010}
Bridge, C. R., et al. 2010,  
\apj, 720, 465
\bibitem[Bunker et al.(2004)]{bunker/etal:2004}
Bunker, A. J., Stanway, E. R., Ellis, R. S., \& McMahon, R. G. 2004
MNRAS, 355, 374
\bibitem[Bunker et al.(2010)]{bunker/etal:2010}
Bunker, A. J., et al. 2010,  
MNRAS, 409, 855
\bibitem[Chen et al.(2007)]{chen:2007}
Chen, H. W., Prochaska, J. X., \& Gnedin, N. Y. 2007,
ApJ, 667, L125

\bibitem[Ciardi et al.(2002)]{ciardi/etal:2002}
Ciardi, B., Bianchi, S., \& Ferrara, A. 2002,
MNRAS, 331, 463

\bibitem[Clark \& Oey(2002)]{clark/oey:2002}
Clarke, C., \& Oey, M. S. 2002,
MNRAS, 337, 1299
\bibitem[Cowie et al.(2009)]{cowie/etal:2009}
Cowie, L. L., Barger, A. J., \& Trouille, L. 2009,
\apj,692,1476
\bibitem[Deharveng, et al.(2001)]{deharveng:2001}
Deharveng, J-M., Buat, V., Le Brun, V., Milliard, B., Kunth, D., Shull, J. M., \& Gry, C. 2001,  
A\&A, 375, 805
\bibitem[Dove \& Shull(1994)]{dove/shull:1994}
Dove, J. B., \& Shull, J. M. 1994,
\apj, 430, 222

\bibitem[Dove et al.(2000)]{dove/shull/ferrara:2000} 
Dove, J. B., Shull, J. M., \& Ferrara, A. 2000,  
\apj, 531, 846 
\bibitem[Dunkley et al.(2009)]{dunkley/etal:2008}
Dunkley, J., et al. 2009, 
ApJS, 180, 306
\bibitem[Ferguson(2001)]{ferguson/etal:2001}
Ferguson, H. C. 2001, in Deep Fields, Proc.\ ESO Workshop, ed. S. Cristiani, 
A. Renzini, R. E. Williams (Garching: Springer-Verlag), 54 

\bibitem[Fernandez \& Komatsu(2006)]{fernandez/komatsu:2006}
Fernandez, E. R., \& Komatsu, E. 2006, 
ApJ, 646, 703
\bibitem[Fernandez-Soto et al.(2003)]{fernandez-soto/etal:2003}
Fernandez-Soto, A., Lanzetta, K. M., \& Chen, H.-W. 2003,
MNRAS, 342, 1215
\bibitem[Fujita et al.(2003)]{fujita/etal:2003}
Fujita, A., Martin, C. L.,  Mac Low, M. M., \& Abel, T. 2003,
\apj, 599, 50
\bibitem[Giallongo, et al.(2002)]{Giallongo/etal:2002}
Giallongo, E., Cristiani, S., D'Odorico, S., \& Fontana, A. 2002,
\apj, 568, L9
\bibitem[Gneiden(2008)]{gnedin:2008}
Gnedin, N. Y. 2008,
\apj, 673, L1

\bibitem[Gnedin et al.(2008)]{gnedin/etal:2008}
Gnedin, N. Y., Kravtsov, A. V., \& Chen, H. -W. 2008,
\apj, 672, 765
\bibitem[Gonzalez et al.(1010)]{gonzalez/etal:2010}
Gonz‡lez, V., LabbŽ, I., Bouwens, R. J., Illingworth, G., Franx, M., Kriek, M., \& Brammer, G. B. 2010,  
\apj, 713, 115
\bibitem[Grimes et al.(2007)]{grimes/etal:2007}
Grimes, J. P., et al. 2007,  
\apj, 668, 891
\bibitem[Grimes et al.(2009)]{grimes:2009}
Grimes, J. P., et al.  2009,  
\apj,181, 272
\bibitem[Hanish et al.(2010)]{hanish/etal:2010}
Hanish, D. J., Oey, M. S., Rigby, J. R., de Mello, D. F., \& Lee, J. C. 2010,
\apj, 725, 2029

\bibitem[Heckman et al.(2001)]{heckman/etal:2001}
Heckman, T. M., Sembach, K. R., Meurer, G. R., Leitherer, C., Calzetti, D., \& Martin, C. L. 2001,  
\apj, 558, 56
\bibitem[Hobson \& Scheuer(1993)]{hobson/scheuer:1993}
Hobson, M. P., \& Scheuer, P. A. 1993,
\mnras, 264, 145
\bibitem[Hoopes et al.(2007)]{hoopes:2007}
Hoopes, C. G., et al. 2007,  
ApJS, 173, 441
\bibitem[Hurwitz et al.(1997)]{hurwitz/etal:1997}
Hurwitz, M., Jelinsky, P., Van Dyke Dixon, W. 1997,
\apj, 481, L31
\bibitem[Inoue et al.(2005)]{inoue/etal:2005}
Inoue, A. K., Iwata, I., Deharveng, J.-M., Buat, V., \& Burgarella, D. 2005
A\&A, 435, 471
\bibitem[Inoue et al.(2006)]{inoue/etal:2006}
Inoue, A. K., Iwata, I., \& Deharveng, J.-M. 2006,
\mnras, 371, L1 
\bibitem[Iwata(2009)]{Iwata/etal:2009}
Iwata, I., et al. 2009,  
\apj, 692, 1287
\bibitem[Kogut et al.(2003)]{kogut/etal:2003}
Kogut, A., et al. 2003, 
\apjs, 148, 161
\bibitem[Komatsu et al.(2009)]{komatsu/etal:2008}
Komatsu, E., et al. 2009, 
ApJS, 180, 330
\bibitem[Komatsu et al.(2010)]{wmap7}
Komatsu, E., et al. 2011, 
ApJS, 192, 18
\bibitem[Larson(1998)]{larson:1998}
Larson, R. B. 1998,
\mnras, 301, 569

\bibitem[Lehnert \& Bremer(2003)]{lehnert/bremer:2003}
Lehnert, M. D., \& Bremer, M. 2003,
\apj, 593, 630

\bibitem[Leitherer et al.(1995)]{Leitherer/etal:1995}
Leitherer, C., Ferguson, H. C., Heckman, T. M., Lowenthan, J. D. 1995,
\apj, 454, L19

\bibitem[Madau et al.(1999)]{madau/etal:1999}
Madau, P., Haardt, F., \& Rees, M. J. 1999, 
\apj, 514, 648 

\bibitem[Malkan, et al.(2003)]{malkan:2003}
Malkan, M., Webb, W., Konopacky, Q. 2002,
\apj, 598, 878
\bibitem[McLure et al.(2010)]{mclure/etal:2010}
McLure, R. J., Dunlop, J. S., Cirasuolo, M., Koekemoer, A. M., Sabbi, E., Stark, D. P., Targett, T. A., \& Ellis, R. S.2010,  
MNRAS, 403, 960
\bibitem[Mo et al.(1998)]{mo/etal:1998}
Mo, H. J., Mao, S., \& White, S. D. 1998,
MNRAS, 295, 319
\bibitem[Navarro et al.(1997)]{navarro/etal:1997}
Navarro, J. F., Frenk, C. S., \& White, S. D. M. 1997,
\apj, 490, 493
\bibitem[Oesch et al.(2010a)]{oesch/etal:2010a}
Oesch, P. A., et al. 2010,  
\apj, 709, L16
\bibitem[Oesch et al.(2010b)]{oesch/etal:2010b}
Oesch, P. A., et al. 2010, 
\apj, 690, 1350
\bibitem[Ouchi et al.(2009)]{ouchi/etal:2009}
Ouchi, M., et al. 2009,  
\apj, 706, 1136

\bibitem[Page et al.(2007)]{page/etal:2007}
 Page, L., et al. 2007,  
 ApJS, 170, 335
\bibitem[Press \& Schechter(1974)]{press/schechter:1974}
Press, W. H., \& Schechter, P. 1974,
\apj, 187, 425
\bibitem[Putman et al.(2003)]{putman/etal:2003}
Putman, M. E., Bland-Hawthorn, J., Veilleux, S., Gibson, B. K., Freeman, K. C., \& Maloney, P. R. 2003 , 
ApJ, 597, 948

\bibitem[Razoumov \& Sommer-Larsen(2006)]{razoumov/sl:2006}
Razoumov, A., \& Sommer-Larsen, J. 2006,
\apj, 651, L89

\bibitem[Razoumov \& Sommer-Larsen(2010)]{razoumov/sl:2009}
Razoumov, A., \& Sommer-Larsen, J. 2010,
\apj, 710, 1239
\bibitem[Richard et al.(2008)]{richard/etal:2008}
Richard, J., Stark, Daniel P., Ellis, Richard S., George, Matthew R., Egami, Eiichi, Kneib, Jean-Paul, Smith, \& Graham P. 2008, 
\apj, 685, 705
\bibitem[Ricotti \& Shull(2000)]{ricotti/shull:2000}
Ricotti, M., \& Shull, J. M. 2000,
\apj, 542, 548

\bibitem[Robertson et al.(2010)]{robertson/etal:2010}
Robertson, B. E., Ellis, R. S., Dunlop, J. S., McLure, R. J., \& Stark, D. P. 2010,
   \nat, 468, 49 


\bibitem[Salpeter(1955)]{salpeter:1955}
Salpeter, E. E. 1955,
\apj, 121, 161
\bibitem[Schaerer(2002)]{schaerer:2002}
Schaerer, D. 2002,
\aap, 382, 28

\bibitem[Shapley et al.(2006)]{shapley/etal:2006}
Shapley, A. E., Steidel, C. C., Pettini, M., Adelberger, K. L., \& Erb, D. K. 2006,
\apj, 651, 688

\bibitem[Siana, et al.(2007)]{Siana/etal:2007}
Siana, B., et al. 2007,  
\apj, 668, 62
\bibitem[Siana et al.(2010)]{siana/etal:2010}
Siana, B., et al. 2010,  
\apj, 723, 241
\bibitem[Spergel et al.(2003)]{spergel/etal:2003}
Spergel, D. N., et al. 2003, ApJS, 148, 175  
\bibitem[Spergel et al.(2007)]{spergel/etal:2007}
Spergel, D. N., et al. 2007, ApJS, 170, 377 
\bibitem[Spitzer(1942)]{spitzer:1942}
Spitzer, L. 1942,
\apj, 95, 329
\bibitem[Srbinovsky \& Wyithe(2008)]{srbinovsky/wyithe:2008}
Srbinovsky, J. A., \& Wyithe, J. S. B. 2008,
PASA, 27, 110
\bibitem[Steidel et al.(2002)]{Steidel:2002}
Steidel, C. C., Pettini, M, \& Adelberger, K. L. 2001,
\apj, 546, 665\bibitem[Stiavelli, Fall, \& Panagia(2004)]{stiavelli/etal:2004}
Stiavelli, M., Fall, S. M., \& Panagia, N. 2004,
\apj, 610, L1
\bibitem[Shull \& Trenti(2010)]{trenti/shull:inprep}
Shull, J. M., \& Trenti, M. 2010, in prep

\bibitem[Trenti et al.(2010)]{trenti/etal:2010}
Trenti, M., Stiavelli, M., Bouwens, R. J., Oesch, P., Shull, J. M., Illingworth, G. D.,
   Bradley, L. D., \& Carollo, C. M. 2010, \apj, 714, 202

\bibitem[Tumlinson et al.(1999)]{tumlinson/etal:1999}
Tumlinson, J., Giroux, M. L., Shull, J. M., \& Stocke, J. T. 1999, \aj, 118, 2148 

\bibitem[Vanzella et al.(2010a)]{vanzella/etal:2010}
Vanzella, E.,  Siana, B., Cristiani, S, \& Nonino, M. 2010,
MNRAS, 404, 1672
\bibitem[Vanzella et al.(2010b)]{vanzella:2010}
Vanzella, E., et al. 2010,  
\apj, 725, 1011

\bibitem[Wise \& Cen(2009)]{wise/cen:2008} 
Wise, J. H., \& Cen, R. 2009, \apj, 693, 984

\bibitem[Witt \& Gordon(1996)]{witt/gordon:1996}
Witt, A. N., \& Gordon, K. D. 1996,
\apj, 463, 681

\bibitem[Wood \& Loeb(2000)]{wood/loeb:2000}
Wood, L., \& Loeb, A. 2000,
ApJ, 545, 86

\bibitem[Wyithe et al.(2010)]{wyithe:2010}
Wyithe, J. S. B., Hopkins, A. M., Kistler, M. D., Yuksel, H., \& Beacom, J. F. 2010,
MNRAS 401, 2561

\bibitem[Yajima et al.(2010)]{yajima/etal:2010}
Yajima, H., Jun-Hwan, C., \& Nagamine, K. 2010,
MNRAS , in press, arXiv:1002.3346


\bibitem[Yan \& Windhorst(2004)]{yan/windhorst:2004}
Yan, H., \& Windhorst, R. A. 2004,
\apj, 612, L93



\end{thebibliography}
\end{document}